# From Group Sparse Coding to Rank Minimization: A Novel Denoising Model for Low-level Image Restoration

Yunyi Li[1], Guan Gui[2,*] and Xiefeng Cheng[1]

[1] College of Electronic and Optical Engineering & College of Microelectronics, Nanjing University of Posts and Telecommunications, Nanjing 210023, China;

[2] College of Telecommunications & Information Engineering, Nanjing University of Posts and Telecommunications, Nanjing 210023, China;

*Corresponding Author: guiguan@njupt.edu.cn

**Abstract:** Recently, low-rank matrix recovery theory has been emerging as a significant progress for various image processing problems. Meanwhile, the group sparse coding (GSC) theory has led to great successes in image restoration (IR) problem with each group contains low-rank property. In this paper, we propose a novel low-rank minimization based denoising model for IR tasks under the perspective of GSC, an important connection between our denoising model and rank minimization problem has been put forward. To overcome the bias problem caused by convex nuclear norm minimization (NNM) for rank approximation, a more generalized and flexible rank relaxation function is employed, namely weighted nonconvex relaxation. Accordingly, an efficient iteratively-reweighted algorithm is proposed to handle the resulting minimization problem combing with the popular $L_{1/2}$ and $L_{2/3}$ thresholding operators. Finally, our proposed denoising model is applied to IR problems via an alternating direction method of multipliers (ADMM) strategy. Typical IR experiments on image compressive sensing (CS), inpainting, deblurring and impulsive noise removal demonstrate that our proposed method can achieve significantly higher PSNR/FSIM values than many relevant state-of-the-art methods.

## I. Introduction

Image restoration (IR) problem [1][2][3] is a key topic in low-level visions. The goal of IR is to reconstruct a high-quality image or sequence $\mathbf{X}$ from its various degraded observations $\mathbf{Y}$. Typical low level IR research includes image compressive sensing (CS) reconstruction [4][5], denoising [6], inpainting [7], deblurring [8] and impulsive noise removal [9], etc. To tackle this typical ill-posed problem, the most popular method is the regularization technique

$$\min_{\mathbf{X}}\{\mathcal{L}(\mathbf{X}) = \mathcal{F}(\mathbf{X},\mathbf{Y}) + \lambda \Re(\mathbf{X})\} \qquad (1)$$

where $\mathcal{F}(\mathbf{X},\mathbf{Y})$ denotes the fidelity term which can penalize our desired image or sequence $\mathbf{X}$ far from the original degraded $\mathbf{Y}$, the second term $\Re(\mathbf{X})$ is the regularization term which can provide the necessary prior knowledge of image, e.g., the sparsity, smoothness or continuity, and the regularization parameter $\lambda$ can make the tradeoff between the first fidelity term and the second regularization term.

It is well documented that how to exploit more prior knowledge for the minimization of problem (1) is at the core, in the past several years, how to design the regularization term to exploit the prior of an image in a predefined domain has been widely studied. Sparsity-inducing approach has been widely exploited in various sparse signal recovery and IR applications [10][11][12][13], e.g. the $L_1$-norm based regularization, the nonconvex penalized regularization of $L_p$-norm [14], Smoothly Clipped Absolute Deviation (SCAD) [15], Logarithm [16], and Minimax Concave Penalty (MCP) [17], utilize the sparsity as prior for minimization. However, these traditional regularization term for IR approaches can only exploit a few structural features of image and some important artifacts will not be preserved. Besides the sparsity based prior model, another popular prior models to exploit the nonlocal self-similarity features of image patches, which can provide more prior knowledge to improve the restoration quality and the promising performance has been well documented [18]. Very recently, the sparsity and the nonlocal self-similarity features often exploit simultaneously to product better approximation [19], one of the state-of-the-art models is the non-locally centralized sparse representation (NCSR) model proposed in [20], which can exploit the nonlocal self-similarity of image to earn a more accurate sparse representation coefficient, and then centralize the sparse coefficients of the degraded image to the restored image, and has shown its promising performance. In [21], a novel and efficient framework is proposed for IR based on group sparse coding (GSC) model, where the image can be sparsely represented as a linear combination in the domain of group.

Although the well-known GSC based model has shown the significant improvements for IR tasks, the important low-rank property of each group is not being used, and still minimizing the optimization problem from the perspective of sparsity. According to the concept of image group matrix, for any given image, the adjacent patches have similar structures, and similar patches can be grouped into a matrix after vectorizing processing, such that the matrix shows low-rank property, then matrix competition for each group can be conducted to recover the desired image. The last decades has seen the group based low-rank minimization method for image and video restoration with the development of the theory of compressive sensing and GSC [22][23][24][25]. Like the $L_0$-norm regularized minimization problem, it is not tractable to minimize the rank regularized problem because of the property of nonconvex and discontinuous of $rank(\mathbf{X})$, and is usually relaxed to the convex nuclear norm minimization (NNM) model. However, the most popular convex NNM often leads to a biased solution, since NNM is usually over-shrink the singular values and treats each of them equally, while larger singular values actually quantify the main preserved information. To tackle this problem, various low-rank relaxations have been proposed, the weighted nuclear norm (WNN) [6], the truncated nuclear norm (TNN) [26], the Schatten-$p$ nuclear norm ($Sp$-NN) [27], and the generalized nonconvex nonsmooth functions on singular values (e.g., $L_p$, SCAD, MCP and Logarithm, e.t,.) [28][29][11].

Recently works tend to exploit the theoretic connection and relationship among various sparsity models and low-rank models. Zha et al. [30][31] answer this important question and build a theoretical benchmark for sparse coding and low-rank minimization from the perspective of GSC. Wen et al. [32] build a more general theoretic relationship between various prior models of sparsity and low-rank model. In this paper, motivated by the simplest and typical denoising model in IR problem, we try to bridge the gap between sparsity prior model and low-rank minimization schemes from a new perspective of denoising model via the famous GSC theory, which will convert the traditional sparse coding denoising problem into the low-rank regularization model. Then a framework is designed for IR applications by combining the alternating direction method of multipliers (ADMM). Our IR framework not only can unify the local sparsity and nonlocal similarity of image simultaneously as image prior, but also can reconstruct image using various matrix competition approaches for rank approximation. Our contributions in this paper can be summarized as:

(1) We propose a novel denoising model base on the GSC framework via rank minimization. In which, we convert the sparsity-inducing optimization problem into the low-rank matrix minimization problem via an effective dictionary learning approach. We also give a theoretical analysis about their connection and relationship between the denoising model and low-rank minimization scheme.

(2) To achieve an accurate approximation of the rank of group matrix for our denoising model, we propose a generalized low-rank minimization model. Our proposed model employs a generalized and flexible weighted scheme and a generalized nonconvex nonsmooth surrogate function on singular values of the group matrix. To solve the generalized rank minimization problem for GSC, we will first convert the optimization model into a GSC based derivative double nonconvex nonsmooth rank (GSC-DNNR) minimization problem, and then develop an iterative reweighted singular-value function thresholding algorithm.

(3) We address the IR problem using our proposed denoising model, an alternative direction method of multipliers (ADMM) framework is integrated, where the sparse coding for GSC based denoising model and the group matrix of desired image will be achieved simultaneously. We further evaluate the proposed framework for four classic IR problems, including image CS reconstruction, inpainting, deblurring, and impulsive noise removal.

The flowchart of our proposed methods for IR problem by rank minimization based GSC denoising model is presented in the Fig. 1.

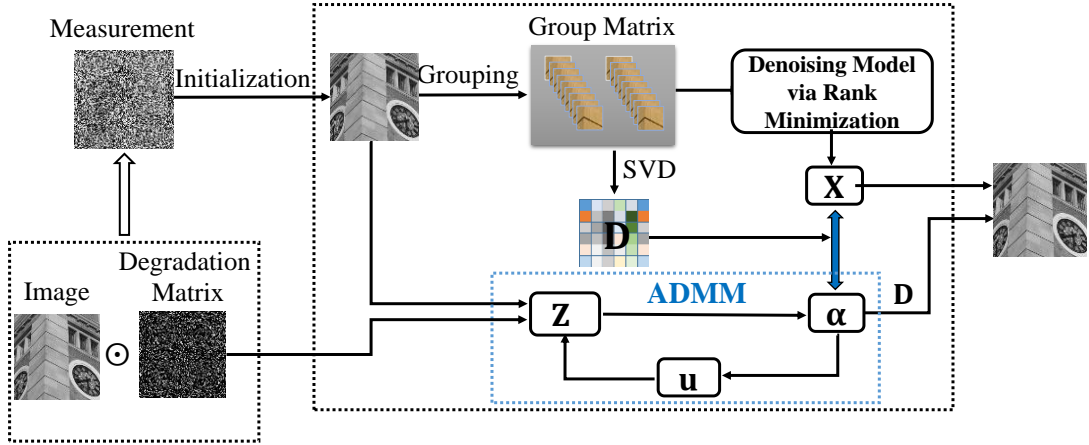

**Fig. 1**. Flowchart of our proposed model for IR problem.

The rest of our paper is as follows. In the section II, we will first introduce the GSC theory, and then propose a GSC based denoising model, and give a theoretical derivation about connection between our proposed denoising model and low-rank minimization problem. Then we exploit a weighted nonconvex relaxation for rank approximation of group matrix and develop an iteratively-reweighted singular-value thresholding algorithm for the GSC based nonconvex nonsmooth low-rank minimization problem. In the section III, we will introduce how to employ our proposed denoising model for IR problems via ADMM framework. In the section IV, we will evaluate the effectiveness of our proposed method for various IR tasks and compare the performance with current state-of-the-art IR algorithms. Finally, a brief summary will be made in section V.

## II. Group Sparse Coding based Denoising Model via Rank Minimization

This section will introduce the basic theory of group sparse coding, and then a self-adaptive dictionary learning strategy is introduced for each group. The GSC problem can be converted into the low-rank matrix recovery problem via our proposed adaptive dictionary learning scheme.

### 2.1. Group sparse coding

Concretely, we first divide the original image $\mathbf{X} \in \mathbb{R}^{\sqrt{N} \times \sqrt{N}}$ into $n$ overlapping patches $\mathbf{X}_k, k = 1, 2, \cdots, n$, with the size is $\sqrt{\mathcal{B}_s} \times \sqrt{\mathcal{B}_s}, \mathcal{B}_s < N$. For each patch $\mathbf{X}_k$, there exist $c$ best matched patches, we denote the set of these patches as $S_{\mathbf{x}_k}$, then we search them in a given searching window with the size of $L \times L$ using the well-known Euclidean distance as the similarity criterion. Next, these similar patches will be stacked into a group matrix with the size of $\mathcal{B}_s \times c$, denoted by $\mathbf{X}_{G_k} = [\mathbf{X}_{G_k,1}, \mathbf{X}_{G_k,2}, \cdots, \mathbf{X}_{G_k,c}] \in \mathbb{R}^{\mathcal{B}_s \times c}$, where each patch will be vectorized as $\mathbf{X}_{G_k,i} \in \mathbb{R}^{\mathcal{B}_s \times 1}, i = 1, 2, \cdots, c$ as the columns. Then the constructed group matrix $\mathbf{X}_{G_k}$ consisting of $c$ patches containing similar structures, the construction process of group can be expressed as

$$\mathbf{X}_{G_k} = G_k(\mathbf{X}), k = 1, 2, \cdots, n. \tag{2}$$

where the operator $G_k(\cdot)$ denotes the group construction from $\mathbf{X}$. **Fig. 2** illustrates the construction process of group.

Conversely, if we average all the group $\mathbf{X}_{G_k}$, we can achieve the original image $\mathbf{X}$ by

$$\mathbf{X} = \sum_{k=1}^n \mathcal{R}_k^T(\mathbf{X}_{G_k}) ./ \sum_{k=1}^n \mathcal{R}_k^T(\mathbf{1}_{\mathcal{B}_s}) \tag{3}$$

where $\mathcal{R}_k^T(\cdot)$ denotes the transpose grouping operator, $\mathbf{1}_{\mathcal{B}_s} \in \mathbb{R}^{\mathcal{B}_s \times c}$ stands for a matrix with all the elements are 1, and operator $./$ denotes an element-wise division of two matrix.

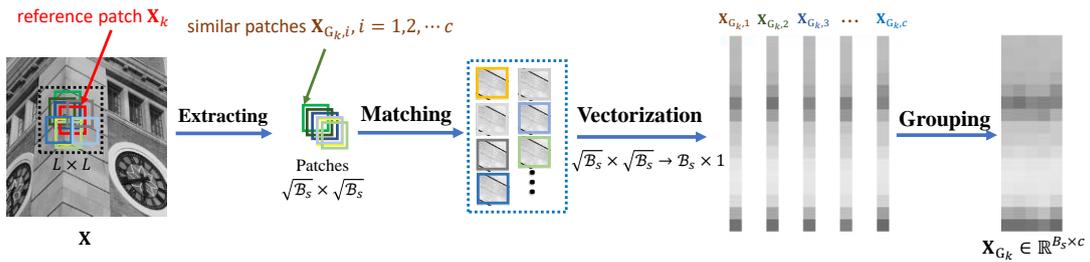

**Fig. 2**. Illustrations of generating group matrix.

According to sparse coding theory, it is expected that the coefficient vector is as sparse as possible with the group $\mathbf{X}_{G_k}$ can be faithfully represented by the dictionary $\mathbf{D}_{G_k}$. To obtain an adaptive dictionary $\mathbf{D}_{G_k}$ for each group $\mathbf{X}_{G_k}$, in this paper, we adopt a self-adaptive dictionary learning scheme [33] for each group, and learn the adaptive dictionary $\mathbf{D}_{G_k}$ from $\mathbf{X}_{G_k}$ directly. We first conduct the singular value decomposition (SVD) of $\mathbf{X}_{G_k}$ by

$$\mathbf{X}_{G_k} = \mathbf{U}_{\mathbf{X}_{G_k}} \mathbf{\Sigma}_{\mathbf{X}_{G_k}} \mathbf{V}_{\mathbf{X}_{G_k}}^T = \sum_{i=1}^m \sigma_{\mathbf{X}_{G_k},i} \boldsymbol{u}_{\mathbf{X}_{G_k},i} \boldsymbol{v}_{\mathbf{X}_{G_k},i}^T \tag{4}$$

where $m = min(B_s, c)$, $\mathbf{U}_{G_k} = [u_{\mathbf{X}_{G_k},1}, u_{\mathbf{X}_{G_k},2}, \cdots, u_{\mathbf{X}_{G_k},m}]$, $\mathbf{V}_{\mathbf{X}_{G_k}} = [v_{\mathbf{X}_{G_k},1}, v_{\mathbf{X}_{G_k},2}, \cdots, v_{\mathbf{X}_{G_k},m}]$ and $\Sigma_{\mathbf{X}_{G_k}} = diag([\sigma_{\mathbf{X}_{G_k},1}; \sigma_{\mathbf{X}_{G_k},2}; \cdots; \sigma_{\mathbf{X}_{G_k},m}])$. Then, the atom $\mathbf{d}_{G_k}$ of the dictionary $\mathbf{D}_{G_k}$ can be obtained by

$$\mathbf{d}_{G_k,i} = u_{\mathbf{X}_{G_k},i} v_{\mathbf{X}_{G_k},i}^{\mathrm{T}}, \quad i = 1, 2, \cdots, m \tag{5}$$

Finally, we can achieve the self-adaptive dictionary for each group by

$$\mathbf{D}_{G_k} = [\mathbf{d}_{G_k,1}, \mathbf{d}_{G_k,2}, \cdots, \mathbf{d}_{G_k,m}]. \tag{6}$$

It should be noted that every atom $\mathbf{d}_{G_k,i} \in \mathbb{R}^{B_s \times c}, i = 1, 2, \cdots, m$ in the dictionary $\mathbf{D}_{G_k}$ is a matrix with the same size of group $\mathbf{X}_{G_k}$. After obtaining the dictionary $\mathbf{D}_{G_k} = [\mathbf{d}_{G_k,1}, \mathbf{d}_{G_k,2}, \cdots, \mathbf{d}_{G_k,m}] \in \mathbb{R}^{(B_s \times c) \times m}$, the representation coefficient vector can be achieved by

$$\hat{\boldsymbol{\alpha}}_{G_k} = \arg\min_{\boldsymbol{\alpha}_{G_k}} \frac{1}{2} \|\mathbf{X}_{G_k} - \mathbf{D}_{G_k} \boldsymbol{\alpha}_{G_k}\|_2^2 + \lambda \|\boldsymbol{\alpha}_{G_k}\|_0, \quad k = 1, 2, \cdots, n. \tag{7}$$

in which $\boldsymbol{\alpha}_{G_k} \in \mathbb{R}^{m \times 1}$ is the achieved sparse coefficient vector. Then the image group will be represented sparsely by

$$\mathbf{X}_{G_k} = \mathbf{D}_{G_k} \hat{\boldsymbol{\alpha}}_{G_k} = \sum_{i=1}^{m} \alpha_{G_k,i} \mathbf{d}_{G_k,i} \tag{8}$$

where $\mathbf{D}_{G_k} = [\mathbf{d}_{G_k,1}, \cdots, \mathbf{d}_{G_k,m}] \in \mathbb{R}^{(B_s \times c) \times m}$ denotes the dictionary, and $\hat{\boldsymbol{\alpha}}_{G_k} = [\hat{\alpha}_{G_k,1}, \hat{\alpha}_{G_k,2}, \cdots, \hat{\alpha}_{G_k,m}] \in \mathbb{R}^{m \times 1}$ is the desired vector.

For simplify, if we concatenate all the corresponding dictionary $\mathbf{D}_{G_k}$ as $\mathbf{D}_G$, and the original entire image $\mathbf{X}$ can be represented by

$$\mathbf{X} = \mathbf{D}_G \circ \boldsymbol{\alpha}_G \tag{9}$$

where $\boldsymbol{\alpha}_G$ denotes the concatenation of all $\mathbf{D}_{G_k}$. As a result, the learned self-adaptive dictionary will build the link between the GSC problem and rank minimization problem.

**2.2. GSC based denoising model via low-rank minimization**

In practice, the original image $\mathbf{X}$ and the corresponding groups $\mathbf{X}_{G_k}, k = 1, 2, \cdots, n.$ are unknown, considering the following GSC based denoising model

$$\hat{\boldsymbol{\alpha}}_{G_k} = \arg\min_{\boldsymbol{\alpha}_{G_k}} \frac{1}{2} \|\mathbf{Y}_{G_k} - \mathbf{D}_{G_k} \boldsymbol{\alpha}_{G_k}\|_2^2 + \lambda \|\boldsymbol{\alpha}_{G_k}\|_0, \quad k = 1, 2, \cdots, n. \tag{10}$$

where $\mathbf{Y}_{G_k} \in \mathbb{R}^{B_s \times c}$ denotes the degraded observation of $\mathbf{X}_{G_k}$, $\boldsymbol{\alpha}_{G_k}$ denotes the sparse coefficient vector over the dictionary $\mathbf{D}_{G_k}$, and $\lambda$ denotes the regularization parameter. Through the GSC model, we can estimate the latent image $\mathbf{X}_{G_k}$ over the adaptive dictionary $\mathbf{D}_{G_k}$ from its degraded observation $\mathbf{Y}_{G_k}$, then the desired matrix (group) can be reconstructed by $\hat{\mathbf{X}}_{G_k} = \mathbf{D}_{G_k} \hat{\boldsymbol{\alpha}}_{G_k}$.

**Remark 2.1.** *For any image group matrix* $\mathbf{X}_{G_k}$ *and its singular value decomposition (SVD)* $\mathbf{X}_{G_k} = \sum_{i=1}^{m} \sigma_{\mathbf{X}_{G_k},i} u_{\mathbf{X}_{G_k},i} v_{\mathbf{X}_{G_k},i}^{\mathrm{T}}$, *if* $\mathbf{X}_{G_k} = \mathbf{D}_{G_k} \boldsymbol{\alpha}_{G_k}$, *then the number of nonzero elements of* $\boldsymbol{\alpha}_{G_k}$ *is equal to the number of nonzero value of singular values* $\sigma_{\mathbf{X}_{G_k},i}$, *we have the following relationship*

$$\|\boldsymbol{\alpha}_{G_k}\|_0 = rank\left(\sum_{i=1}^{m} \sigma_{\mathbf{X}_{G_k},i} \mathbf{d}_{G_k,i}\right) = rank(\mathbf{D}_{G_k} \boldsymbol{\alpha}_{G_k}) = rank(\mathbf{X}_{G_k}) \tag{11}$$

where $\boldsymbol{\alpha}_{G_k} = [\alpha_{G_k,1}, \alpha_{G_k,2}, \cdots, \alpha_{G_k,m}] \in \mathbb{R}^{m \times 1}$, $\mathbf{D}_{G_k} = [\mathbf{d}_{G_k,1}, \cdots, \mathbf{d}_{G_k,m}] \in \mathbb{R}^{(B_s \times c) \times m}$, and $rank(\mathbf{X}_{G_k})$ *denotes the singular value number of the matrix* $\mathbf{X}_{G_k}$.

Benefiting from the fact that the degraded group $\mathbf{Y}_{G_k}$ and original group $\mathbf{X}_{G_k}$ share the same coding dictionary $\mathbf{D}_{G_k}$, according to the relationship in the **Remark 2.1,** by substituting $\mathbf{D}_{G_k}\boldsymbol{\alpha}_{G_k}$ using $\mathbf{X}_{G_k}$ in (10), the denoising problem (10) can be converted into the following equivalent low-rank minimization

$$\widehat{\mathbf{X}}_{G_k} = \arg\min_{\mathbf{X}_{G_k}} \frac{1}{2}\|\mathbf{Y}_{G_k} - \mathbf{X}_{G_k}\|_F^2 + \lambda rank(\mathbf{X}_{G_k}), k = 1, 2, \cdots, n. \tag{12}$$

where $\mathbf{X}_{G_k}$ denotes the constructed image group with low-rank property, and $\mathbf{Y}_{G_k}$ denotes the corresponding observation group. It is worth noting that the sparse coefficient vector $\boldsymbol{\alpha}_{G_k}$ is also the singular value vector of matrix $\mathbf{X}_{G_k}$ in our proposed model. Then we can convert the sparsity-inducing optimization problem (10) into the low-rank minimization problem (12).

The low-rank minimization problem (12) is a NP-hard problem, which is usually relaxed as a convex NNM problem,

$$\widehat{\mathbf{X}}_{G_k} = \arg\min_{\mathbf{X}_{G_k}} \frac{1}{2}\|\mathbf{Y}_{G_k} - \mathbf{X}_{G_k}\|_F^2 + \lambda\|\mathbf{X}_{G_k}\|_* \tag{13}$$

where, $\|\mathbf{X}_{G_k}\|_* = \sum_{i=1}^{min(B_s,c)} |\sigma_i(\mathbf{X}_{G_k})|$ denotes the nuclear norm, and $\sigma_i(\mathbf{X}_{G_k}), i = 1,2,\cdots,$ are the singular values of $\mathbf{X}_{G_k}$, and $\|\cdot\|_F$ presents the Frobenius norm. Many recent works tend to use the improved NNM for low-rank minimization, e.g., the truncated nuclear norm (TNN) [26], the Schatten $p$-norm (S$p$) [27], the weighted nuclear norm (WNN) [6], and so on. It is well documented that the WNN based method can often earn the best performance among them [6]. What's more, some popular nonconvex counterparts of $L_0$-norm on singular value have shown great potentials to improve the rank minimization performance [28][29], typical nonconvex surrogate functions include $L_p$-norm [14], SCAD [15], Logarithm [16], MCP [17]. Motivated by the weighted strategy and the nonconvex counterparts of $L_0$-norm, this paper uses a more generalized and flexible rank relaxation function for rank approximation of the group matrix using weighted nonconvex relaxation, such that

$$\widehat{\mathbf{X}}_{G_k} = \arg\min_{\mathbf{X}_{G_k}} \frac{1}{2}\|\mathbf{Y}_{G_k} - \mathbf{X}_{G_k}\|_F^2 + \lambda\boldsymbol{\rho}_{w_k}\left(\sigma(\mathbf{X}_{G_k})\right) \tag{14}$$

in which, $\boldsymbol{\rho}_{w_k}\left(\sigma(\mathbf{X}_{G_k})\right) = \sum_{i=1}^{r} w_{k,i}\rho(\sigma_{k,i})$, $r = min(B_s,c)$, $\boldsymbol{w}_k = (w_{k,1}, w_{k,2}, \cdots, w_{k,r})$ is the weighting vector with $w_{k,1} \leq w_{k,2} \leq \cdots \leq w_{k,r}$, $\sigma(\mathbf{X}_{G_k}) = (\sigma_{k,1}, \sigma_{k,2}, \cdots, \sigma_{k,r})$ denotes the singular value vector with $\sigma_{k,1} \leq \sigma_{k,2} \leq \cdots \leq \sigma_{k,r}$, the function $\rho(\cdot): \mathbb{R}^+ \to \mathbb{R}^+$ is the proper and lower semi-continuous function, and is nondecreasing on $[0, +\infty)$. It should be noted that the nonconvexity of the function $\rho(\cdot)$ is often weaker than traditional nonconvex functions, e.g., $L_p$-norm, MCP and SCAD. According to the definition of the rank relaxation function, $\boldsymbol{\rho}_w(\cdot)$ will be more flexible with different $w_i$ and $\rho(\cdot)$. When $\rho(\cdot)$ is the absolute function, $\boldsymbol{\rho}_w(\cdot)$ becomes the traditional nuclear norm and the weighted nuclear norm with all $w_i = 1$ and not all $w_{k,i} = 1$, respectively. When $\rho(\cdot)$ is the $L_p$-norm with $0 < p < 1$, $\boldsymbol{\rho}_w(\cdot)$ will become the Schatten $p$-nuclear norm and the weighted Schatten $p$-nuclear norm with all $w_{k,i} = 1$ and not all $w_{k,i} = 1$, respectively. Moreover, when the weighting vector $\boldsymbol{w}$ with partial $w_{k,i} = 0$, $\boldsymbol{\rho}_w(\cdot)$ will become the truncated nuclear norm and the truncated Schatten p-nuclear norm with $p = 1$ and $0 < p < 1$, respectively. A summarization for these special cases can be found in the **Table** 1. Our proposed model can estimate the rank of each group with a high accurate and nearly unbiased solution, one popular example is the weighting vector $\boldsymbol{w}_k$ is inversely proportional to the corresponding singular values, e.g., $\omega_{k,i} = 1/(|\sigma_i(\mathbf{X}_{G_k})| + \varepsilon)$ [6].

**Table 1**. Different relaxation functions with different weight $w$ and relaxation function $\rho$.

| Name | Relaxation Function | $w$ | $\rho$ | $\partial\rho$ |
|---|---|---|---|---|
| Nuclear norm | $\|\mathbf{X}\|_* = \sum_{i=1}^m \sigma_i(\mathbf{X})$ | $w_i = 1, i = 1,2,\cdots,m$ | $\rho(\sigma_i) = \sigma_i,$ $(i=1,\cdots,m)$ | $\rho(\sigma_i) = \sigma_i,$ $(i=1,\cdots,m)$ |
| Weighted nuclear norm | $\|\mathbf{X}\|_{w,*} = \sum_{i=1}^m w_i \sigma_i(\mathbf{X})$ | $w_i \geq 0, i = 1,2,\cdots,m$ | | |
| Truncated nuclear norm | $\|\mathbf{X}\|_r = \sum_{i=r+1}^m \sigma_i(\mathbf{X})$ | $w_i = 1, i = r+1,\cdots,m$ | $\rho(\sigma_i) = \begin{cases} 0, i = 1,\cdots,r \\ \sigma_i, i = r+1,\cdots,m \end{cases}$ | $\rho(\sigma_i) = \begin{cases} 0, i = 1,\cdots,r \\ 1, i = r+1,\cdots,m \end{cases}$ |
| Weighted Truncated nuclear norm | $\|\mathbf{X}\|_{w,r} = \sum_{i=r+1}^m w_i \sigma_i(\mathbf{X})$ | $w_i \geq 0, i = r+1,\cdots,m$ | | |
| Schatten $p$-norm | $\|\mathbf{X}\|_{Sp}^p = \sum_{i=1}^m (\sigma_i(\mathbf{X}))^p$ | $w_i = 1, i = 1,2,\cdots,m$ | $\rho(\sigma_i) = (\sigma_i)^p,$ $(i=1,\cdots,m)$ | $\rho(\sigma_i) = p(\sigma_i)^{p-1},$ $(i=1,\cdots,m)$ |
| Weighted Schatten $p$-norm | $\|\mathbf{X}\|_{w,Sp}^p = \sum_{i=1}^m w_i (\sigma_i(\mathbf{X}))^p$ | $w_i \geq 0, i = 1,2,\cdots,m$ | | |

Actually, according to the reweighted strategy [28] and the super-gradient properties [28], the problem (14) can be intrinsically derived from the following nonconvex nonsmooth rank minimization problem [34][35], e.g.,

$$\widehat{\mathbf{X}}_{G_k} = \arg\min_{\mathbf{X}_{G_k}} \frac{1}{2} \|\mathbf{Y}_{G_k} - \mathbf{X}_{G_k}\|_F^2 + \lambda \sum_{i=1}^r h\left(\rho\left(\sigma_i(\mathbf{X}_{G_k})\right)\right) \quad (15)$$

where $h(\rho(\cdot))$ denotes the relaxation function, and in this paper, we choose $h(\cdot) = \rho(\cdot)$ without loss of the generality, in which, $\rho(\cdot)$ denotes a function with lower semi-continuous property in $[0, +\infty)$. As a result, the problem (15) can be substituted by

$$\widehat{\mathbf{X}}_{G_k} = \arg\min_{\mathbf{X}_{G_k}} \frac{1}{2} \|\mathbf{Y}_{G_k} - \mathbf{X}_{G_k}\|_F^2 + \lambda \sum_{i=1}^r \rho\left(\rho\left(\sigma_i(\mathbf{X}_{G_k})\right)\right). \quad (16)$$

in which, there exist two relaxation function $\rho(\cdot)$, thus (16) can be known as the GSC based double nonconvex nonsmooth rank (GSC-DNNR) minimization problem.

### 2.3. GSC-DNNR minimization for denoising model

The GSC-DNNR minimization problem (16) is more difficult to resolve than traditional convex NNM problem. To solve this problem, we usually need to convert the minimization problem into the iteratively reweighted minimization problem, which benefits from the antimonotone property of relaxation function $\rho(\cdot)$ and the supergradient function $\partial\rho(\cdot)$. We first give the following theorem.

**Lemma 2.1.** [28] *Let $\rho(\cdot): \mathbb{R}^+ \to \mathbb{R}^+$ be concave with the monotonical nondecreasing property, and its supergradient $\partial\rho(\cdot)$ is monotonically nonincreasing on $[0, +\infty)$, then we have*

$$\rho(s_1) - \rho(s_2) \leq w_{s_2}(s_1 - s_2) \quad (17)$$

*where $w_{s_2} \in \partial\rho(s_2)$, $s_1$ and $s_2$ denote two any given values.*

Since the monotonical nondecreasing property of $\rho(\cdot)$ and the monotonical nonincreasing property of $\partial\rho(\cdot)$, for ang given $\sigma_1 \geq \sigma_2 \geq \cdots \geq \sigma_r$, we have $\rho(\sigma_1) \geq \rho(\sigma_2) \geq \cdots \geq \rho(\sigma_r)$ and $\partial\rho(\rho(\sigma_1)) \leq \partial\rho(\rho(\sigma_2)) \leq \cdots \leq \partial\rho(\rho(\sigma_r))$, that is to say, the function of $\partial\rho(\rho(\cdot))$ is monotonical nonincreasing on $[0, +\infty)$. Then according to the **Lemma 2.1**, if we replace the value $s_1$ and $s_2$ by $\rho(s_1)$ and $\rho(s_2)$, and substrate the weighting $w_{s_2} \in \partial\rho(s_2)$ by $w_{\rho(s_2)} \in \partial\rho(\rho(s_2))$. Then we have the following theorem.

**Lemma 2.2.** [35] *Let $\rho(\cdot): \mathbb{R}^n \to \mathbb{R}$ be concave with the monotonical nondecreasing property, its supergradient $\partial\rho(\cdot)$ is monotonically nonincreasing on $[0, +\infty)$, and the double supergradient $\partial\rho(\rho(\cdot))$ is monotonically nonincreasing on $[0, +\infty)$, then we have*

$$\rho(\rho(s_1)) - \rho(\rho(s_2)) \leq w_{\rho(s_2)}(\rho(s_1) - \rho(s_2)) \quad (18)$$

*where $w_{\rho(s_2)} \in \partial\rho(\rho(s_2))$, $s_1$ and $s_2$ denote two any given values.*

According to the **Lemma 2.2**, since the relaxation function of $\rho(\cdot)$ is concave on $[0, +\infty)$, then according to the monotonical nondecreasing property of double supergradient $\partial\rho(\rho(\cdot))$. As a result, we have

$$\rho\left(\rho\left(\sigma_i(\mathbf{X}_{G_k})\right)\right) \leq \rho\left(\rho\left(\sigma_i(X_{G_k}^t)\right)\right) + w_{k,i}^t \left(\rho\left(\sigma_i(\mathbf{X}_{G_k})\right) - \rho\left(\sigma_i(\mathbf{X}_{G_k}^t)\right)\right) \quad (19)$$

where $w_{k,i}^t \in \partial\rho\left(\rho\left(\sigma_i(X_{G_k}^t)\right)\right)$ with $w_{k,1}^t \leq w_{k,2}^t \leq \cdots \leq w_{k,r}^t$. Then the problem (16) can be relaxed as the following optimization problem, e.g.,

$$\mathbf{X}_{G_k}^{t+1} = \arg\min_{\mathbf{X}_{G_k}} \frac{1}{2}\|\mathbf{Y}_{G_k} - \mathbf{X}_{G_k}\|_F^2 + \lambda \sum_{i=1}^r \left\{\rho\left(\rho\left(\sigma_i(\mathbf{X}_{G_k}^t)\right)\right) + w_{k,i}^t \left(\rho\left(\sigma_i(\mathbf{X}_{G_k})\right) - \rho\left(\sigma_i(\mathbf{X}_{G_k}^t)\right)\right)\right\}$$

$$= \arg\min_{\mathbf{X}_{G_k}} \frac{1}{2}\|\mathbf{Y}_{G_k} - \mathbf{X}_{G_k}\|_F^2 + \lambda \sum_{i=1}^r w_{k,i}^t \rho\left(\sigma_i(\mathbf{X}_{G_k})\right) \quad (20)$$

in which, the iteratively-weighting vector can be updated by $w_{k,i}^{t+1} \in \partial\rho\left(\rho\left(\sigma_i(\mathbf{X}_{G_k}^{t+1})\right)\right)$, and $\mathbf{X}_{G_k}^t$ denotes the $k$-th iteration of variable $\mathbf{X}_{G_k}$.

Essentially, to solve (20) is equivalent to solving a proximal operator. Some previous works have been studied to solve the (20) for some special cases of the relaxation functions, such as the iteratively reweighted nuclear norm (IRNN) algorithm (Lu et, al. [28]). Here, we will develop our algorithm for GSC proposed in Zhang et, al [35], which can be a extend version of Lu's IRNN method (Lu et, al. [28]). We will present that, if the singular-value function thresholding operator is monotone, the problem (20) can be resolved by the weighted operator.

**Theorem 2.1.** [29] *Let a function $\rho(\cdot): \mathbb{R}^n \to \mathbb{R}$ such that a proximal operator $\text{Prox}_{w_i,\rho}^{\delta_i}(\cdot)$ is monotone, i.e., $\text{Prox}_{w_1,\rho}^{\delta_1}(\delta_1) > \text{Prox}_{w_2,\rho}^{\delta_2}(\delta_2)$ for any $\delta_1 > \delta_2$. For any given $\lambda > 0$, $\mathbf{Y} \in \mathbb{R}^{m \times n}$, $r = \min(m,n)$, the weighting values with $w_1 \leq w_2 \leq \cdots \leq w_r$. Let $\mathbf{Y} = \mathbf{U}Diag(\delta(\mathbf{Y}))\mathbf{V}^\mathrm{T}$ is the SVD of $\mathbf{Y}$, and the singular values satisfy $\delta_1(\mathbf{Y}) \geq \delta_2(\mathbf{Y}) \geq \cdots \geq \delta_r(\mathbf{Y})$, then the solution to*

$$\mathbf{X} = \arg\min_{\mathbf{X}} \frac{\gamma}{2}\|\mathbf{X} - \mathbf{Y}\|_F^2 + \lambda \sum_{i=1}^r w_i \rho(\sigma_i(\mathbf{X})) \quad (21)$$

*is*

$$\mathbf{X}^* = \mathbf{U}Diag(\delta_i^*)\mathbf{V}^\mathrm{T} \quad (22)$$

*where $\delta_i^*$ can be obtained by*

$$\delta_i^* \in \text{Prox}_{w_i,\rho}^{\delta_i}(\delta_i(\mathbf{Y})) = \arg\min_{\sigma_i > 0} \frac{1}{2}(\sigma_i(\mathbf{X}) - \delta_i(\mathbf{Y}))^2 + \lambda w_i \rho(\sigma_i(\mathbf{X})) \quad (23)$$

It follows from the **Theorem 2.1.**, considering the problem (20), since $w_{k,1}^t \leq w_{k,2}^t \leq \cdots \leq w_{k,r}^t$, and $\rho(\cdot): \mathbb{R}^n \to \mathbb{R}$ and the proximal operator $\text{Prox}_{w_i,\rho}^{\delta_i}(\cdot)$ is monotone. Then the optimal solution of our optimization problem (20) can be eventually achieved by

$$\mathbf{X}_{G_k}^{t+1} = \mathbf{U}Diag\left(\text{Prox}_{w_k,\rho}^{\delta}\left(\delta(\mathbf{Y}_{G_k}^t)\right)\right)\mathbf{V}^\mathrm{T} \quad (24)$$

where $\text{Prox}^{\delta}_{w_k,\rho}\left(\delta(\mathbf{Y}^t_{G_k})\right)$ denotes the element-wise operator, e.g.,

$$\text{Prox}^{\delta_i}_{w_{k,i},\rho}\left(\delta_i(\mathbf{Y}^t_{G_k})\right) = \arg\min_{\delta_i>0}\frac{1}{2}\left(\sigma_i(\mathbf{X}_{G_k}) - \delta_i(\mathbf{Y}^t_{G_k})\right)^2 + \lambda w^t_{k,i}\rho\left(\sigma_i(\mathbf{X}_{G_k})\right) \quad (25)$$

where $\delta_1(\mathbf{Y}^t_{G_k}) \geq \delta_2(\mathbf{Y}^t_{G_k}) \geq \cdots \geq \delta_r(\mathbf{Y}^t_{G_k})$.

## 2.4. $L_{1/2}$ and $L_{2/3}$ thresholding operator

We next give the closed-form solution of (25). It should be noted that the problem (25) is a weighted version with a nonconvex relaxation function $\rho(\cdot)$. In this paper, we employ the popular $L_p$-function with $p = 1/2$ and $p = 2/3$ as the relaxation surrogates. The reason is that the $L_p$-function is more flexible, moreover, two special cases of $p = 1/2$ and $p = 2/3$ have be demonstrated their high efficiency and can earn their closed-form solutions [36].

(1) $L_p$-norm with $p = 1/2$, then the problem (25) is reduced to

$$\sigma^*_i(\mathbf{X}_{G_k}) = \arg\min_{\delta_i>0}\frac{1}{2}\left(\sigma_i(\mathbf{X}_{G_k}) - \delta_i(\mathbf{Y}^t_{G_k})\right)^2 + \xi_i\left(\sigma_i(\mathbf{X}_{G_k})\right)^{1/2} \quad (26)$$

then the closed-form of (25) can be defined by [37]

$$\sigma^*_i(\mathbf{X}_{G_k}) = \begin{cases} \frac{2}{3}\sigma_i(\mathbf{X}_{G_k})\left(1 + \cos\left(\frac{2\pi}{3} - \frac{2}{3}\varphi\left(\sigma_i(\mathbf{X}_{G_k})\right)\right)\right), & |\sigma_i(\mathbf{X}_{G_k})| > T \\ 0, & \text{otherwise} \end{cases} \quad (27)$$

where $\varphi\left(\sigma_i(\mathbf{X}_{G_k})\right) = \cos^{-1}\left(\frac{\xi_i}{4}\left(\frac{|\sigma_i(\mathbf{X}_{G_k})|}{3}\right)^{-3/2}\right)$, $\xi_i = (\lambda w^t_{k,i})$, and the threshold value $T = \frac{3\sqrt[3]{2}}{4}(2\xi_i)^{2/3}$.

(2) $L_p$-norm with $p = 2/3$, then the problem (25) is reduced to

$$\sigma^*_i(\mathbf{X}_{G_k}) = \arg\min_{\delta_i>0}\frac{1}{2}\left(\sigma_i(\mathbf{X}_{G_k}) - \delta_i(\mathbf{Y}^t_{G_k})\right)^2 + \xi_i\left(\sigma_i(\mathbf{X}_{G_k})\right)^{2/3} \quad (28)$$

then the closed-form of (25) can be defined by [36]

$$\sigma^*_i(\mathbf{X}_{G_k}) = \begin{cases} \left(\left(|\theta(\sigma_i(\mathbf{X}_{G_k}))| + \sqrt{2|\sigma_i(\mathbf{X}_{G_k})|/|\theta(\sigma_i(\mathbf{X}_{G_k}))| - |\theta(\sigma_i(\mathbf{X}_{G_k}))|^2}\right)/2\right)^3, & |\sigma_i(\mathbf{X}_{G_k})| > T \\ 0, & \text{otherwise} \end{cases} \quad (29)$$

where $\theta\left(\sigma_i(\mathbf{X}_{G_k})\right) = \frac{2}{\sqrt{3}}(2\xi_i)^{1/4}\left(\cosh\left(\frac{1}{3}\text{arccosh}\left(\frac{27}{16}(2\xi_i)^{-3/2}\left(\sigma_i(\mathbf{X}_{G_k})\right)^2\right)\right)\right)^{1/2}$, $\xi_i = (\lambda w^t_{k,i})$ and $T = \frac{2\sqrt[4]{3}}{3}(2\xi_i)^{3/4}$ denotes the threshold value.

The whole algorithm for GSC can be summarized in the **Algorithm 1**.

---

**Algorithm 1**: Proposed GSC-DNNR algorithm for GSC based denoising optimization model

$$\widehat{\mathbf{X}}_{G_k} = \arg\min_{\mathbf{X}_{G_k}}\frac{1}{2}\|\mathbf{Y}_{G_k} - \mathbf{X}_{G_k}\|^2_F + \lambda\rho_{w_k}\left(\sigma(\mathbf{X}_{G_k})\right)$$

**Input:** The Observation $\mathbf{Y}$;

**Initialization:** $t = 0, \gamma > L_g, \mathbf{X}_{G_k}^0, \mathbf{w}_k^0$;
**For** $t = 1, 2, \cdots$ **do**
**While not converge do**
    1. Updating $\mathbf{X}_{G_k}^{t+1}, \boldsymbol{\alpha}_{G_k}^{t+1}$ by solving (16);

    2. Updating the weighting vector $\mathbf{w}_k^t$ by $w_{k,i}^{t+1} \in \partial \rho \left( \rho \left( \sigma_i \left( X_{G_k}^{t+1} \right) \right) \right)$;

**End**
**Output:** $\widehat{\mathbf{X}}_{G_k}, \boldsymbol{\alpha}_{G_k}^{t+1}$.

## III. Integrating GSC based Denoising Model to Image Restoration Applications via ADMM

In this section, we will integrate the GSC based denoising model to IR problems. Considering the following image degraded model

$$\mathbf{b} = \mathbf{A}\mathbf{x} + \mathbf{n} \tag{30}$$

where $\mathbf{b} \in \mathbb{R}^{\sqrt{M} \times \sqrt{N}}$ denotes the degraded observation, $\mathbf{A} \in \mathbb{R}^{\sqrt{M} \times \sqrt{N}}$ denotes the degradation matrix, $\mathbf{x} \in \mathbb{R}^{\sqrt{N} \times \sqrt{N}}$ and $\mathbf{n}$ are the desired image and the additive noise. Different $\mathbf{A}$ will cause different IR task, when $\mathbf{A}$ is a compressed sampling operator, the IR problem becomes compressive sensing [38], and an identity matrix of $\mathbf{A}$ with entries are either 1 or 0 will often cause image inpainting problem [7]. Suppose the original image $\mathbf{X} \in \mathbb{R}^N$ (also $\mathbf{X} \in \mathbb{R}^{\sqrt{N} \times \sqrt{N}}$) can be represented by the sparse coefficient vector in the domain $\boldsymbol{\Psi}$, denotes as $\boldsymbol{\alpha} = \boldsymbol{\Psi}\mathbf{X}$, that is $\mathbf{X} = \mathbf{D}\boldsymbol{\alpha}$, where $\mathbf{D}$ denotes the corresponding dictionary, which can be known or learned from the images, and the IR problem can be described as the following generalized optimization model

$$\widehat{\boldsymbol{\alpha}} = \arg\min_{\boldsymbol{\alpha}} \mathcal{F}(\mathbf{b}; \boldsymbol{\alpha}) + \lambda \mathfrak{R}(\boldsymbol{\alpha}) \tag{31}$$

where $\mathcal{F}(\mathbf{b}; \boldsymbol{\alpha})$ is the fidelity term and $\mathfrak{R}(\boldsymbol{\alpha})$ denotes regularization term, which measures the sparsity degree of the image and can provide prior knowledge for minimization, such as $\|\boldsymbol{\alpha}\|_p$, the parameter $\lambda$ denotes the regularization parameter. Then our desired image can be reconstructed by $\widehat{\mathbf{X}} = \mathbf{D}\widehat{\boldsymbol{\alpha}}$.

### 3.1 ADMM framework for image restoration via nonconvex weighted group sparse coding

According to the alternative direction method of multipliers (ADMM) framework, we introduce an auxiliary variable $\mathbf{z}$ to the problem (31),

$$\widehat{\boldsymbol{\alpha}} = \arg\min_{\boldsymbol{\alpha}, \mathbf{z}} \mathcal{F}(\mathbf{b}; \mathbf{z}) + \lambda \mathfrak{R}(\boldsymbol{\alpha}), \ s.t. \ \mathbf{z} = \mathbf{D}\boldsymbol{\alpha} \tag{32}$$

Without confusion, we have the following iterative steps:

$$\begin{cases} \mathbf{z}^{(t+1)} = \arg\min_{\mathbf{z}} \mathcal{F}(\mathbf{b}; \mathbf{z}) + \frac{\mu}{2} \left\| \mathbf{z} - \mathbf{D}\boldsymbol{\alpha}^{(t)} - \mathbf{u}^{(t)} \right\|_2^2 \\ \boldsymbol{\alpha}^{(t+1)} = \arg\min_{\boldsymbol{\alpha}} \frac{\mu}{2} \left\| \mathbf{z}^{(t+1)} - \mathbf{D}\boldsymbol{\alpha} - \mathbf{u}^{(t)} \right\|_2^2 + \lambda \mathfrak{R}(\boldsymbol{\alpha}) \\ \mathbf{u}^{(t+1)} = \mathbf{u}^{(t)} - \left( \mathbf{z}^{(t+1)} - \mathbf{D}\boldsymbol{\alpha}^{(t+1)} \right) \end{cases} \tag{33}$$

Then our optimization problem can be split into two subproblems of $\mathbf{z}$ and $\boldsymbol{\alpha}$.

#### 3.1.1 z-subproblem

The $\mathbf{z}$-subproblem is also a simplified reconstruction step which depends on the forward model, this paper utilizes the very popular quadratic function to fit the data, e.g., $\mathcal{F}(\mathbf{b};\mathbf{z}) = \frac{1}{2}\|\mathbf{b} - \mathbf{A}\mathbf{z}\|_2^2$, leading to a quadratic problem, which has a closed-form solution expressed as

$$\mathbf{z} = (\mathbf{A}^T\mathbf{A} + \mu\mathbf{I})^{-1}(\mathbf{A}^T\mathbf{b} + \mu(\mathbf{D}\boldsymbol{\alpha} + \mathbf{u})) \tag{34}$$

where $\mathbf{I}$ denotes the identity matrix. It is efficient to achieve the solution by (34) for image inpainting problem without computing the matrix inverse because of the specific structure in observation matrix $\mathbf{A}$. However, it will be too time-consuming for CS reconstruction problem. To avoid the computing of matrix inverse, here, a gradient descent method is adopted to solve the $\mathbf{z}$-subproblem for CS reconstruction by

$$\tilde{\mathbf{z}} = \mathbf{z} - \eta\mathbf{d} \tag{35}$$

where the parameter $\eta$ denotes the optimal step size, and $\mathbf{d}$ denotes the gradient direction of $\frac{1}{2}\|\mathbf{b} - \mathbf{A}\mathbf{z}\|_2^2 + \frac{\mu}{2}\|\mathbf{z} - \mathbf{D}\boldsymbol{\alpha} - \mathbf{u}\|_2^2$, and we have

$$\mathbf{d} = \mathbf{A}^T\mathbf{A}\mathbf{z} - \mathbf{A}^T\mathbf{b} + \mu(\mathbf{z} - \mathbf{D}\boldsymbol{\alpha} - \mathbf{u}). \tag{36}$$

### 3.1.2 Denoising operator

After achieving the $\mathbf{z}$, the $\boldsymbol{\alpha}$ subproblem can be expressed as

$$\boldsymbol{\alpha}^{(t+1)} = \arg\min_{\boldsymbol{\alpha}} \frac{\mu}{2}\|\mathbf{R}^{(t+1)} - \mathbf{D}\boldsymbol{\alpha}\|_2^2 + \lambda\mathfrak{R}(\boldsymbol{\alpha}) \tag{37}$$

where $\mathbf{R}^{(t+1)} = \mathbf{z}^{(t+1)} - \mathbf{u}^{(t)}$ can be regarded as the degraded observation of $\mathbf{z}^{(t+1)}$. It should be noted that the optimization problem (38) is a typical denoising model. To improve the performance for IR problem, we can generate the corresponding group matrix by grouping the similar patches, e.g., $\mathbf{R}_{G_k}$, $\mathbf{D}_{G_k}$ and $\boldsymbol{\alpha}_{G_k}$. As a result, we have the following equivalent equation with the probability (near to 1) according to following **Theorem 3.1**, e.g.,

$$\|\mathbf{R} - \mathbf{X}\|_2^2 = \frac{N}{K}\sum_{k=1}^{n}\|\mathbf{R}_{G_k} - \mathfrak{R}(\mathbf{X}_{G_k})\|_F^2 \tag{38}$$

where $\mathbf{X} = \mathbf{D}\boldsymbol{\alpha}$.

**Theorem 3.1** [39] *Assume* $\mathbf{R}, \mathbf{X} \in \mathbb{R}^{\sqrt{N}\times\sqrt{N}}, \mathbf{R}_{G_k}, \mathbf{X}_{G_k} \in \mathbb{R}^{B_s\times c}$, *we define* $\mathbf{E} = \mathbf{R} - \mathbf{X}$ *as the error matrix, and its each element is* $\mathbf{E}(j), j = 1, \cdots \sqrt{N} \times \sqrt{N}$, *If each element* $\mathbf{E}(j)$ *is independent and the corresponding distribution of* $\mathbf{E}$ *with zero mean and variance* $\delta^2$, *Then for any* $\varepsilon > 0$, *we have the following relationship between* $\|\mathbf{R} - \mathbf{X}\|_2^2$ *and* $\|\mathbf{R}_{G_k} - \mathbf{X}_{G_k}\|_F^2$, *that is*

$$\lim_{\substack{N\to\infty \\ K\to\infty}} P\left\{\left|\frac{1}{N}\|\mathbf{R} - \mathbf{X}\|_2^2 = \frac{1}{K}\sum_{k=1}^{n}\|\mathbf{R}_{G_k} - \mathbf{X}_{G_k}\|_F^2\right| < \varepsilon\right\} = 1 \tag{39}$$

Then optimization problem (37) can be written as

$$\boldsymbol{\alpha}^{(t+1)} = \arg\min_{\boldsymbol{\alpha}_{G_k}} \frac{1}{2}\sum_{k=1}^{m}\|\mathbf{R}_{G_k} - \mathbf{D}_{G_k}\boldsymbol{\alpha}_{G_k}\|_2^2 + \tau\sum_{k=1}^{m}\mathfrak{R}(\boldsymbol{\alpha}_{G_k}) \tag{40}$$

where $\tau = \frac{\lambda K}{\mu N}$ with $K = n \times c \times B_s$. According to the theory described in the section 2.2, and by substituting $\mathbf{X}_{G_k} = \mathbf{D}_{G_k}\boldsymbol{\alpha}_{G_k}$, then the problem (38) can be transformed into the following GSC framework based optimization problem,

$$\mathbf{X}^{(t+1)} = \arg\min_{\mathbf{X}_{G_k}} \frac{1}{2}\sum_{k=1}^{m}\|\mathbf{X}_{G_k} - \mathbf{R}_{G_k}\|_F^2 + \tau \sum_{k=1}^{m} \Re(\mathbf{X}_{G_k}) \qquad (41)$$

where $\mathbf{X}_{G_k} \in \mathbb{R}^{B_s \times c}$ denotes the group matrix with low-rank property. As mentioned before, the denoising model only depends on the prior term, after combining our proposed generalized rank minimization based GSC model, the optimization (41) can be converted into the low-rank matrix recovery problem, that is

$$\mathbf{X}^{(t+1)} = \arg\min_{\mathbf{X}_{G_k}} \frac{1}{2}\sum_{k=1}^{m}\|\mathbf{X}_{G_k} - \mathbf{R}_{G_k}\|_F^2 + \tau \sum_{k=1}^{m} \rho_{w_k}\left(\sigma(\mathbf{X}_{G_k})\right) \qquad (42)$$

where $\rho_{w_k}\left(\sigma(\mathbf{X}_{G_k})\right) = \sum_{i=1}^{r} \rho\left(\rho\left(\sigma_i(\mathbf{X}_{G_k})\right)\right)$ denotes the proposed relaxation function. As a result, the optimization problem (42) can be split into $m$ subproblems, and the $k, (k = 1,2,\cdots,m)$-*th* subproblem can be shown as

$$\mathbf{X}_{G_k}^{(t+1)} = \arg\min_{\mathbf{X}_{G_k}} \frac{1}{2}\|\mathbf{X}_{G_k} - \mathbf{R}_{G_k}\|_F^2 + \tau \sum_{i=1}^{r} \rho\left(\rho\left(\sigma_i(\mathbf{X}_{G_k})\right)\right). \qquad (43)$$

With this, the optimization problem (43) can be solved by our proposed algorithm in **Algorithm 1**. It should be noted that problem (43) and (16) are two same minimization problems exactly with different regularization parameters $\tau$, therefore, the $\boldsymbol{\alpha}$-subproblem in IR can be resolved via our proposed nonconvex weighted GSC model.

Then we can achieve the closed-form solution of (43) by **Algorithm 1**, i.e.,

$$\mathbf{X}_{G_k}^{(t+1)} = \mathbf{U}_{R_{G_k}} \text{Prox}_\rho^\delta\left(\delta(\mathbf{R}_{G_k}^t)\right) \mathbf{V}_{R_{G_k}}^T, \; i = 1,\cdots,min(B_s,c) \qquad (44)$$

where $\mathbf{R}_{G_k} = \mathbf{U}_{R_{G_k}} \mathbf{\Sigma}_{R_{G_k}} \mathbf{V}_{R_{G_k}}^T$ denotes its singular value decomposition (SVD), $\mathbf{\Sigma}_{G_k} = diag\left(\xi_{R_{G_k},1},\cdots,\xi_{R_{G_k},min(B_s,c)}\right)$ and the operator $(\theta)_+ = max(\theta,0)$. It should be noted that the sparse coefficient vector $\boldsymbol{\alpha}_{G_k}$ in problem (10) is the singular value vector of matrix $\mathbf{X}_{G_k}$ in our proposed model (see subsection 2.3), hence we can achieve the $\boldsymbol{\alpha}_{G_k}^{t+1}$ simultaneously from (44).

### 3.2. Summary and Discussion of the proposed algorithm

We first give a whole summary about the proposed algorithm. When all the groups $\{\mathbf{X}_{G_k}\}, k = 1,2,\cdots,n$ are known, then the latent image $\mathbf{X}$ can be reconstructed by

$$\mathbf{X} = \sum_{k=1}^{n} \mathcal{R}_k^T(\mathbf{X}_{G_k}) ./ \sum_{k=1}^{n} \mathcal{R}_k^T(\mathbf{1}_{B_s \times c}) \qquad (45)$$

where the $\mathcal{R}_k^T(\cdot)$ denotes the transpose grouping operator, $\mathbf{1}_{\mathcal{B}_c}$ denotes a column matrix with the size of $\mathcal{B}_c \times c$ and all elements being 1, the operator $./$ denotes an element-wise division of two matrix. The proposed whole algorithm of nonconvex weighted GSC for IR via ADMM can be summarized as the **Algorithm 2**.

According to our proposed **Algorithm 2**, we can easily find that our proposed method earns a very important feature of plug-and-play AMMM [40]. Firstly, the denoising model play a key role, which only depends on the image prior model. Secondly, it is very convenient and flexible to employ some other state-of-the art denoising models in our proposed algorithm. From this perspective, an interesting future work is to exploit advanced plug-and-play algorithm to improve the IR performance.

| **Algorithm 2**: Proposed algorithm for IR problem via ADMM |
|---|
| **Input:** The Observation $\mathbf{Y}$, the degradation matrix $\mathbf{H}$; |

**Initialization:** $c$, $\mathcal{B}_c$, $t = 0$, $\lambda, \mu$, $\gamma$, $\mathbf{u}$, $\mathbf{z}^{(0)}$,
for
    **if** $\mathbf{A}$ is a mask operator
        Updating $\mathbf{z}$ using the Eq. (34);
    **elseif** $\mathbf{A}$ is a blur operator
        Updating $\mathbf{z}$ using the Eq. (34);
    **else** $\mathbf{A}$ is a random projection operator
        Updating $\mathbf{z}$ using the Eq. (35) and (36);
    **end**
    Computing $\mathbf{R}^{(t+1)} = \mathbf{z}^{(t+1)} - \mathbf{u}^{(t)}$;
    Constructing the groups $\{\mathbf{R}_{G_k}\}$;
    **for each group** $\mathbf{R}_{G_k}$
        Construct adaptive dictionary $\mathbf{D}_{G_k}$ using Eq. (4), (5) and (6);
        Reconstruct $\mathbf{X}_{G_k}^{(t+1)}$ and computing $\boldsymbol{\alpha}_{G_k}^{t+1}$ ;
    **end for**
    Updating $\mathbf{D}^{(t+1)}$ by concatenating all $\mathbf{D}_{G_k}$;
    Updating $\boldsymbol{\alpha}^{(t+1)}$ by concatenating all $\boldsymbol{\alpha}_{G_k}$;
    Computing $\mathbf{u}^{(t+1)}$;
    Computing $\mathbf{X}^{(t+1)}$ by concatenating all the dictionaries $\{\mathbf{X}_{G_k}\}$;
    $t = t + 1$;
end
Output: The reconstructed image $\mathbf{X}$.

## IV. Experimental Results

In this section, we will employ the typical $L_p$, ($p = 1/2, 2/3$) as relaxation of nuclear norm to evaluate the effectiveness of our proposed method for typical IR tasks compared with several state-of-the-art competing IR algorithms. We also analyze the convergence of our proposed algorithm on various IR problems. To measure the reconstruction performance quantitatively, two popular metrics of PSNR and feature similarity (FSIM) [41] will be calculated. The widely used test images are employed to evaluate our proposed method presented in Fig. 3. All the best results will be highlighted in bold in tables, and all the simulation experiments are conducted in a personal computer with Intel (R) Core (TM) i7-6770HQ @ 2.6GHz CPU with 16 GB memory and a Windows 10 operating system.

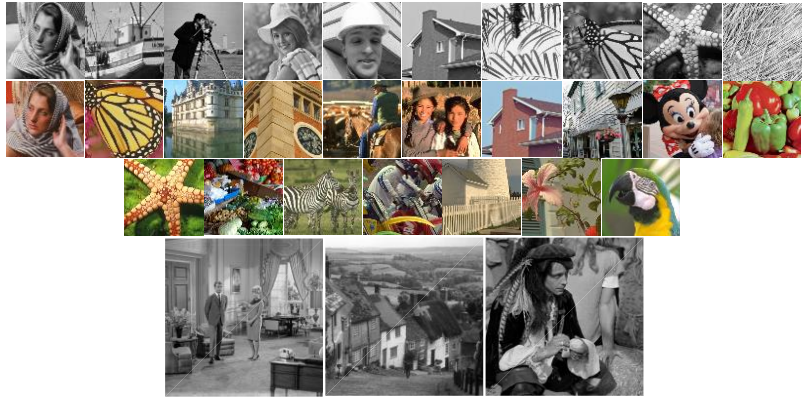

**Fig. 3**. Experimental image set. (a), Typical $256 \times 256$ gray natural images: Barbara, Boats, Cameraman, Elain, Foreman, House, Leaves, Monarch, Starfish and Straw. (b), Typical $256 \times 256$ color images. Top: Barbara, Butterfly, Castle, Clock, Cowboy, Girl, House, Light; Bottom: Mickey, Peepers, Starfish, Vegetable, Zebra, Bike, Fence, Flower and Parrots. (c), Typical $512 \times 512$ gray natural images: Couple, Hill and Man.

### 4.1 Compressive Sensing

CS based image reconstruction technology aims to capture high-quality images from a small number of under-sampling random measurements, in which, one of the main technical challenges is how to obtain high-quality images while reducing the number of measurements. To evaluate the validation of our proposed method, six representative competing CS reconstruction algorithms are employed for comparisons, including the algorithms of BCS [42], SGSR [43], ALSB [39], JASR [44], GSR-Lp [45], GSR-NCR [46]. We empirically set the similar patch parameter $c = 60$, the patch size of $\sqrt{\mathcal{B}_s} \times \sqrt{\mathcal{B}_s}$ is selected as $6 \times 6$, the block size is $32 \times 32$, and the searching window of $L \times L$ is set to be $20 \times 20$, and $\varepsilon = 0.1$ for all of our CS experiments. The other parameter selections of $(\mu, \lambda)$ are listed in the **Table 2** for different sub-sampling rate and surrogate functions. Table 3 lists all achieved PSNR and FSIM results of four competing algorithms and our proposed method under different sampling rates of 0.1, 0.2, 0.3 and 0.4. We can observe that our proposed rank minimization based GSC method can obtain higher PSNR values and FSIM values than all competing methods. To make a visual comparison, we present the reconstructed results of boats and monarch from 0.1 measurements using our proposed method and other approaches, shown in the Fig. 4 and 5. These visual results demonstrate that our proposed method can reconstruct image with higher quality.

**Table 2.**  The parameter selection for five penalty functions under different sub-rates

| Different penalties | 0.10 | 0.20 | 0.30 | 0.40 |
|---|---|---|---|---|
| $p = \frac{1}{2}$ $(\mu, \lambda)$ | $(8e-3, 0.10)$ | $(3e-2, 0.30)$ | $(5e-3, 0.01)$ | $(2.5e-2, 0.08)$ |
| $p = \frac{2}{3}$ $(\mu, \lambda)$ | $(9e-3, 0.04)$ | $(9e-3, 0.01)$ | $(5e-3, 0.005)$ | $(2.5e-2, 0.01)$ |

**Table 3.**  The PSNR (dB)/ FSIM achieved by our proposed algorithm and four competing algorithms

| Subrate = 0.100 | | | | | | | | | | | |
|---|---|---|---|---|---|---|---|---|---|---|---|
| **Method** | *Barbara* | *Boats* | *C.man* | *Elaine* | *Foreman* | *House* | *Leaves* | *Monarch* | *Starfish* | *Straw* | Averaged |
| **BCS** | 22.80/0.7891 | 24.52/0.8029 | 21.60/0.7605 | 27.46/0.8811 | 29.76/0.8911 | 26.90/0.8455 | 18.54/0.6852 | 21.70/0.7828 | 22.71/0.8049 | 19.10/0.6559 | 23.51/0.7899 |
| **SGSR** | 28.70/0.9147 | 27.71/0.8915 | 22.60/0.8065 | 31.32/0.9220 | 34.88/0.9393 | 32.77/0.9187 | 22.22/0.8356 | 24.27/0.8371 | 22.91/0.8177 | 20.24/0.7931 | 26.76/0.8676 |

| Method | Barbara | Boats | C.man | Elaine | Foreman | House | Leaves | Monarch | Starfish | Straw | Averaged |
|---|---|---|---|---|---|---|---|---|---|---|---|
| ALSB | 27.01/0.8903 | 27.75/0.8830 | 23.29/0.8057 | 30.99/0.9184 | 33.49/0.9254 | 32.18/0.9069 | 21.37/0.7934 | 24.27/0.8218 | 23.63/0.8343 | 20.61/0.7910 | 26.46/0.8570 |
| JASR | 29.58/0.9223 | 28.59/0.9035 | 23.54/0.8139 | 32.01/0.9282 | 35.61/0.9437 | 33.49/0.9167 | 23.62/0.8799 | 25.83/0.8822 | 24.39/0.8516 | 21.02/0.7970 | 27.77/0.8839 |
| GSR-Lp | 28.38/0.9062 | 28.37/0.8983 | **24.78/0.8412** | 31.27/0.9229 | 35.57/0.9473 | 33.46/0.9269 | 25.17/0.9064 | 26.61/0.9003 | 24.96/0.8649 | 21.09/0.7606 | 27.97/0.8875 |
| GSR-NCR | 28.28/0.9217 | 27.62/0.8977 | 22.50/0.8006 | 31.35/0.9318 | 35.59/0.9449 | 32.35/0.9128 | 21.74/0.8367 | 23.86/0.8289 | 22.92/0.8227 | 20.14/0.7904 | 26.64/0.8688 |
| Proposed ($L_{1/2}$) | **29.99/0.9291** | **29.04/0.9088** | 24.49/0.8421 | **32.39/0.9330** | **36.08/0.9509** | **34.07/0.9301** | 25.82/0.9157 | **27.32/0.9071** | 25.34/0.8758 | **21.49/0.8121** | **28.60/0.9005** |
| Proposed ($L_{2/3}$) | 29.77/0.9246 | 29.12/0.9086 | 24.59/0.8339 | 32.20/0.9296 | 35.95/0.9491 | 34.02/0.9276 | **25.93/0.9149** | 27.34/0.9065 | **25.58/0.8783** | 21.57/0.8067 | 28.61/0.8980 |

Subrate = 0.200

| Method | Barbara | Boats | C.man | Elaine | Foreman | House | Leaves | Monarch | Starfish | Straw | Averaged |
|---|---|---|---|---|---|---|---|---|---|---|---|
| BCS | 24.31/0.8429 | 27.05/0.8640 | 24.65/0.8357 | 31.19/0.9280 | 32.88/0.9296 | 30.58/0.9014 | 21.24/0.7567 | 25.21/0.8465 | 25.27/0.8616 | 20.70/0.7611 | 26.31/0.8528 |
| SGSR | 33.45/0.9615 | 32.41/0.9465 | 26.53/0.8850 | 34.86/0.9551 | 36.98/0.9598 | 35.81/0.9502 | 28.74/0.9373 | 28.76/0.9132 | 27.19/0.8993 | 24.51/0.8854 | 30.92/0.9293 |
| ALSB | 31.77/0.9501 | 33.04/0.9512 | 26.53/0.8794 | 35.11/0.9597 | 35.33/0.9460 | 35.93/0.9541 | 27.14/0.9094 | 28.39/0.8965 | 27.20/0.8973 | 23.88/0.8794 | 30.43/0.9223 |
| JASR | 34.16/0.9651 | 33.21/0.9521 | 27.75/0.8961 | 35.66/0.9603 | 37.87/0.9636 | 36.10/0.9425 | 30.24/0.9516 | 30.60/0.9409 | 29.10/0.9295 | 24.95/0.8913 | 31.96/0.9393 |
| GSR-Lp | 33.74/0.9627 | 33.34/0.9550 | **28.47/0.9096** | 35.72/0.9628 | 38.65/0.9702 | 37.02/0.9630 | 30.33/0.9569 | 31.04/0.9453 | 29.01/0.9312 | 24.73/0.8829 | 32.21/0.9440 |
| GSR-NCR | 33.91/0.9642 | 33.30/0.9526 | 26.30/0.8796 | 35.61/0.9600 | 37.74/0.9578 | 36.57/0.9508 | 28.89/0.9415 | 29.41/0.9201 | 27.88/0.9158 | 24.41/0.8846 | 31.40/0.9327 |
| Proposed ($L_{1/2}$) | **34.62/0.9674** | 34.03/0.9591 | 28.38/0.9086 | **36.15/0.9647** | 38.79/0.9699 | 37.18/0.9632 | **31.44/0.9619** | 31.79/0.9503 | **29.96/0.9409** | **25.12/0.8940** | **32.75/0.9480** |
| Proposed ($L_{2/3}$) | 34.63/0.9676 | **34.09/0.9596** | 27.91/0.9051 | 36.05/0.9650 | **38.61/0.9699** | **37.18/0.9656** | 31.36/0.9621 | **31.63/0.9482** | 29.80/0.9400 | 25.00/0.8933 | 32.63/0.9476 |

Subrate = 0.300

| Method | Barbara | Boats | C.man | Elaine | Foreman | House | Leaves | Monarch | Starfish | Straw | Averaged |
|---|---|---|---|---|---|---|---|---|---|---|---|
| BCS | 25.70/0.8780 | 28.93/0.8995 | 27.12/0.8798 | 33.70/0.9512 | 35.16/0.9504 | 32.87/0.9298 | 23.31/0.8062 | 27.70/0.8839 | 27.17/0.8954 | 22.24/0.8288 | 28.39/0.8903 |
| SGSR | 35.91/0.9762 | 35.22/0.9684 | 28.81/0.9224 | 36.87/0.9695 | 38.47/0.9711 | 37.37/0.9648 | 32.98/0.9676 | 31.99/0.9469 | 30.79/0.9447 | 27.33/0.8288 | 33.57/0.9563 |
| ALSB | 34.70/0.9716 | 36.45/0.9744 | 28.96/0.9148 | 37.49/0.9742 | 36.50/0.9575 | 38.36/0.9730 | 31.30/0.9537 | 31.37/0.9296 | 30.43/0.9412 | 26.05/0.9152 | 33.16/0.9505 |
| JASR | 36.59/0.9785 | 36.08/0.9723 | 29.93/0.9311 | 36.83/0.9661 | 38.54/0.9649 | 38.04/0.9649 | 33.70/0.9719 | 33.63/0.9610 | 32.33/0.9580 | 27.88/0.9366 | 34.36/0.9605 |
| GSR-Lp | 35.67/0.9749 | 35.30/0.9699 | 30.02/0.9329 | 37.39/0.9731 | 40.34/0.9791 | 38.32/0.9729 | 33.17/0.9725 | 33.39/0.9619 | 31.59/0.9540 | 26.72/0.9197 | 34.19/0.9611 |
| GSR-NCR | 37.16/0.9815 | **37.26/0.9783** | 29.37/0.9305 | 38.25/0.9774 | **41.18/**0.9829 | 39.37/0.9795 | 34.92/0.9799 | 34.64/0.9666 | 33.17/0.9654 | 27.56/0.9348 | 35.29/0.9677 |
| Proposed ($L_{1/2}$) | 37.13/0.9812 | 37.17/0.9776 | 29.88/0.**9379** | 38.13/0.9769 | 41.15/0.9822 | **39.42/0.9795** | **35.23/0.9808** | 34.75/0.9663 | **33.30/0.9661** | 27.71/0.9365 | 35.39/0.9685 |
| Proposed ($L_{2/3}$) | **37.18/0.9811** | 37.11/0.9772 | **30.06/0.9370** | 38.20/0.9769 | 41.11/0.9819 | 39.38/0.9787 | 35.17/0.9804 | **34.81/0.9670** | 33.26/0.9657 | **27.89/0.9384** | **35.42/0.9684** |

Subrate = 0.400

| Method | Barbara | Boats | C.man | Elaine | Foreman | House | Leaves | Monarch | Starfish | Straw | Averaged |
|---|---|---|---|---|---|---|---|---|---|---|---|
| BCS | 27.19/0.9069 | 30.60/0.9246 | 29.13/0.9094 | 35.70/0.9656 | 37.04/0.9646 | 34.67/0.9491 | 25.23/0.8459 | 29.75/0.9115 | 28.97/0.9213 | 23.77/0.8750 | 30.21/0.9174 |
| SGSR | 37.70/0.9835 | 37.41/0.9793 | 30.64/0.9465 | 38.63/0.9784 | 39.84/0.9871 | 38.99/0.9759 | 35.83/0.9799 | 34.66/0.9648 | 33.66/0.9661 | 29.53/0.9563 | 35.69/0.9710 |
| ALSB | 37.23/0.9830 | 38.88/0.9838 | 31.11/0.9441 | 39.48/0.9830 | 42.62/0.9871 | 40.06/0.9820 | 34.47/0.9730 | 34.52/0.9581 | 33.24/0.9642 | 28.24/0.9468 | 35.99/0.9705 |
| JASR | 37.39/0.9803 | 37.19/0.9764 | 31.81/0.9519 | 38.28/0.9742 | 41.19/0.9808 | 38.79/0.9676 | 36.56/0.9831 | 36.15/0.9739 | 34.30/0.9677 | 30.05/0.9601 | 36.17/0.9716 |
| GSR-Lp | 38.33/0.9854 | 38.43/0.9832 | 32.07/0.9563 | 39.59/0.9828 | 42.42/0.9866 | 40.56/0.9834 | 36.86/0.9860 | 36.47/0.9767 | 34.73/0.9735 | 29.34/0.9539 | 36.88/0.9768 |
| GSR-NCR | **39.22/0.9879** | **39.63/0.9867** | 31.59/0.9558 | 40.08/0.**9848** | 42.95/0.**9886** | 41.11/0.**9862** | 38.51/0.9894 | 37.58/0.9794 | 36.21/0.**9794** | 30.05/0.9608 | 37.69/0.9799 |
| Proposed ($L_{1/2}$) | 39.18/0.9875 | 39.54/0.9861 | **32.09/0.9590** | **40.06/0.9844** | **43.09/0.9881** | **41.13/0.9855** | 38.61/0.**9895** | **37.62/0.9794** | **36.15/0.9791** | **30.20/0.9617** | **37.77/0.9800** |
| Proposed ($L_{2/3}$) | 39.08/0.9872 | 39.45/0.9857 | 32.05/0.9587 | 39.95/0.9841 | 43.10/0.9883 | 41.13/0.9859 | **38.67/0.9859** | 37.46/0.9782 | 36.11/0.9790 | 30.05/0.9608 | 37.71/0.9798 |

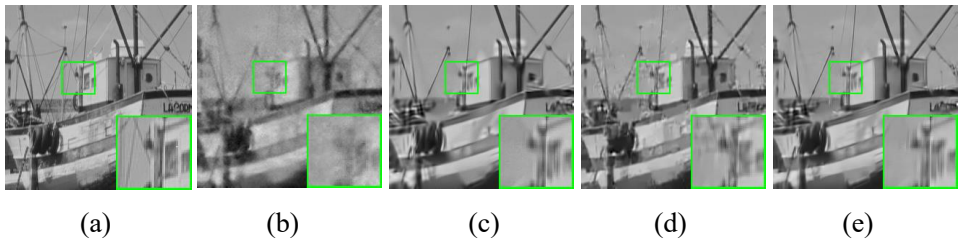

(a)　　　　(b)　　　　(c)　　　　(d)　　　　(e)

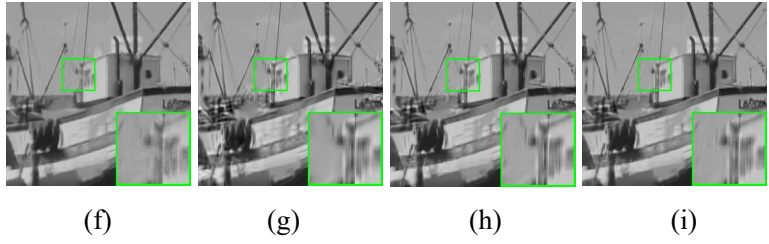

| (f) | (g) | (h) | (i) |

**Fig. 4**. Original image and eight reconstructed images by BCS, SGSR ALSB, ALSB, JASR and our proposed algorithms for 0.1 measurements of boats. (a) Original Image; (b) BCS, 24.52 dB, FSIM=0.8029; (c) SGSR, PSNR=27.71 dB, FSIM=0.8915; (d) ALSB, PSNR=27.75 dB, FSIM=0.8830; (e) JASR, PSNR=28.59 dB, FSIM=0.9035; (f) GSR-Lp, PSNR=28.37 dB, FSIM=0.8983; (g) GSR-NCR, PSNR=27.62 dB, FSIM=0.8977; (h) Proposed ( $p = 1/2$ ), PSNR=**29.04** dB, FSIM=**0.9088**; (i) Proposed ($p = 2/3$), PSNR=**29.12** dB, FSIM=**0.9086**;

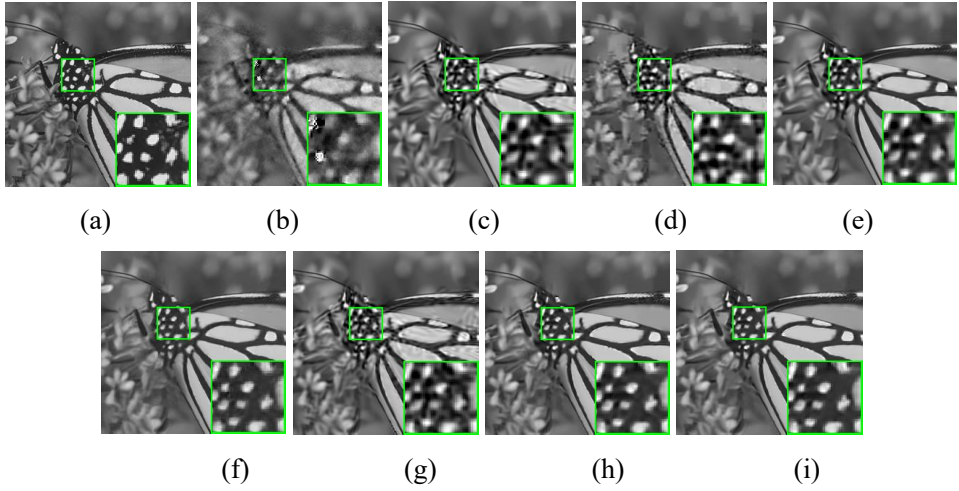

| (a) | (b) | (c) | (d) | (e) |
| (f) | (g) | (h) | (i) |

**Fig. 5**. Visual comparison of the original image and five reconstructed images by BCS, ALSB, ALSB, GSR, JASR and our proposed algorithms for 0.1 measurements of monarch. (a) Original Image; (b) BCS, PSNR=21.70 dB, FSIM=0.7828; (c) SGSR, PSNR=24.27 dB, FSIM=0.8371; (d)ALSB, PSNR=24.27 dB, FSIM=0.8218; (e) JASR, PSNR=25.83 dB, FSIM=0.8822; (f) GSR-Lp, PSNR=26.61 dB, FSIM=0.9003; (g) GSR-NCR, PSNR=23.86 dB, FSIM=0.8289; (h) Proposed ( $p = 1/2$ ), PSNR=**27.32** dB, FSIM=**0.9071**; (i) Proposed ( $p = 2/3$ ), PSNR=**27.34** dB, FSIM=**0.9065**.

### 4.2 Image Inpainting

For the second applications, we employ our proposed nonconvex framework for image inpainting problem. Image inpainting aims at recovering the missing or damaged pixels in images in a plausible way. In this paper, we focus on two interesting cases of restoration from text removal and partial random samples. Here, the size of each patch is $\sqrt{\mathcal{B}_s} \times \sqrt{\mathcal{B}_s} = 8 \times 8$ and $10 \times 10$ for partial random sample and text removal, respectively. The similar patches number $c$ is set to be 60, and the searching window of $L \times L$ is set to be $20 \times 20$, and $\varepsilon = 0.1$ for all image inpainting experiments, and other parameters selection are listed in the **Table 4**.

**Table 4.** The parameter selection for different sub-rate and functions

| Different penalties | 80% | 70% | 50% | Embed text |
|---|---|---|---|---|
| $p=\frac{1}{2}$ $(\mu,\lambda)$ | $(5e-3, 0.1)$ | $(1e-3, 0.025)$ | $(1e-3, 0.01)$ | $(1e-3, 0.01)$ |
| $p=\frac{2}{3}$ $(\mu,\lambda)$ | $(1.6e-1, 0.06)$ | $(1e-1, 0.05)$ | $(3e-1, 0.0007)$ | $(1e-2, 0.005)$ |

## I. Image restoration from partial random samples

We first handle the IR problem from partial random samples, where the degraded matrix **A** is generated by a random matrix. We perform experiments by randomly removing 50%, 70% or 80% of pixels in original images. We employ four state-of-the-art image inpainting algorithms for comparisons, including the algorithms of BPFA [47] (an effective sparse image representations method via Bayesian dictionary learning), IPPO [48] (an IR method based on smooth ordering of patches), Aloha [49] (an recent image inpainting approach based on a low-rank Hankel matrix Approach) and JSM [50] (an IR method using joint statistical modeling scheme). TNNM [33] (a truncated singular value thresholding operator based on the GSC framework), and WNNM [6] (a state-of-the art NNM method). Here, we employ eight color images for experiments. The results achieved by proposed algorithm and other competing state-of-the art algorithms are listed in the **Table 5**, our proposed approach can outperform other competitive methods significantly.

**Table 5.** The PSNR (dB)/FSIM values achieved by our algorithm and other four competing algorithms

| | Missing pixels = 80% | | | | | | | | | | |
|---|---|---|---|---|---|---|---|---|---|---|---|
| Method | *Butterfly* | *Bike* | *Castle* | *House* | *Girl* | *Light* | *Mickey* | *Parrots* | *Vegetable* | *Zebra* | Averaged |
| BPFA | 24.04/0.8533 | 22.81/0.8469 | 23.94/0.8639 | 30.16/0.8987 | 24.80/0.8766 | 19.39/0.8091 | 24.53/0.8696 | 27.28/0.9285 | 23.26/0.8672 | 20.99/0.8190 | 24.12/0.8633 |
| IPPO | 25.13/0.9078 | 24.04/0.8777 | 24.50/0.8818 | 33.65/0.9488 | 25.31/0.8914 | 21.51/0.8757 | 26.33/0.9099 | 29.16/0.9435 | 23.07/0.8619 | 22.71/0.8665 | 25.54/0.8965 |
| Aloha | 24.88/0.8586 | 23.63/0.8675 | 23.89/0.8729 | 33.80/0.9467 | 25.16/0.8832 | 21.49/0.8644 | 25.33/0.8770 | 29.85/0.9474 | 23.04/0.8704 | 22.74/0.8565 | 25.38/0.8845 |
| JSM | 25.58/0.9125 | 23.56/0.8670 | 24.57/0.8829 | 34.28/0.9467 | 25.18/0.8871 | 20.20/0.8524 | 24.57/0.9060 | 28.46/0.9372 | 23.29/0.8665 | 21.86/0.8520 | 25.16/0.8910 |
| TNNM | 26.38/0.9238 | 24.43/0.8855 | 24.94/0.8852 | 35.34/0.9580 | 25.72/0.8954 | 22.41/0.8914 | 26.59/0.9112 | 29.97/0.9485 | 23.43/0.8657 | 22.68/0.8645 | 26.19/0.9029 |
| WNNM | 26.70/0.9291 | 24.88/0.9029 | 24.91/0.8959 | 35.59/0.9592 | 26.05/0.9094 | 21.88/0.8936 | 26.96/0.9212 | 30.09/0.9517 | 23.68/0.8866 | 22.85/0.8816 | 26.36/0.9131 |
| Proposed ($L_{1/2}$) | 26.66/0.9289 | 24.87/0.9023 | 25.03/0.8990 | 35.65/0.9596 | 26.01/0.9107 | 22.72/0.9086 | 27.09/0.9234 | 30.54/0.9509 | 23.65/0.8884 | 23.24/0.8902 | 26.55/0.9162 |
| Proposed ($L_{2/3}$) | 26.39/0.9217 | 24.45/0.8941 | 24.66/0.8899 | 35.15/0.9429 | 25.72/0.9015 | 22.07/0.9036 | 26.85/0.9162 | 29.97/0.9424 | 23.46/0.8813 | 22.63/0.8817 | 26.14/0.9075 |
| | Missing pixels = 70% | | | | | | | | | | |
| Method | *Butterfly* | *Bike* | *Castle* | *House* | *Girl* | *Light* | *Mickey* | *Parrots* | *Vegetable* | *Zebra* | Averaged |
| BPFA | 26.52/0.8985 | 24.85/0.8917 | 25.68/0.9059 | 33.96/0.9447 | 26.61/0.9168 | 21.36/0.8724 | 26.17/0.9024 | 29.91/0.9508 | 24.66/0.9046 | 22.73/0.8706 | 26.25/0.9058 |
| IPPO | 27.68/0.9400 | 26.03/0.9204 | 26.11/0.9162 | 36.64/0.9695 | 27.43/0.9316 | 23.47/0.9173 | 28.59/0.9406 | 32.20/0.9629 | 24.80/0.9070 | 24.76/0.9148 | 27.77/0.9320 |
| Aloha | 27.33/0.8998 | 25.60/0.9130 | 25.84/0.9103 | 36.76/0.9691 | 27.12/0.9215 | 23.19/0.9066 | 27.10/0.9104 | 32.19/0.9641 | 24.52/0.9099 | 24.54/0.9003 | 27.42/0.9205 |
| JSM | 27.95/0.9434 | 25.68/0.9162 | 26.63/0.9227 | 36.82/0.9672 | 27.21/0.9277 | 23.17/0.9189 | 28.22/0.9357 | 31.53/0.9600 | 24.81/0.9083 | 23.97/0.9035 | 27.60/0.9304 |
| TNNM | 28.58/0.9452 | 25.90/0.9125 | 26.55/0.9122 | 36.90/0.9652 | 27.35/0.9217 | 23.93/0.9198 | 28.31/0.9315 | 32.17/0.9576 | 24.63/0.8951 | 24.70/0.9032 | 27.90/0.9264 |
| WNNM | 28.78/0.9506 | 26.39/0.9245 | 26.74/0.9194 | 36.90/0.9686 | 27.50/0.9297 | 24.20/0.9283 | 28.70/0.9389 | 33.05/0.9633 | 24.96/0.9078 | 24.93/0.9117 | 28.22/0.9343 |
| Proposed ($L_{1/2}$) | 29.19/0.9530 | 27.03/0.9362 | 26.99/0.9269 | 37.28/0.9712 | 27.98/0.9384 | 24.45/0.9342 | 29.19/0.9448 | 33.71/0.9676 | 25.28/0.9181 | 25.30/0.9222 | 28.64/0.9413 |
| Proposed ($L_{2/3}$) | 29.10/0.9493 | 26.80/0.9326 | 26.74/0.9253 | 37.15/0.9632 | 28.02/0.9374 | 24.11/0.9335 | 29.12/0.9428 | 33.37/0.9612 | 25.18/0.9183 | 24.91/0.9227 | 28.45/0.9386 |
| | Missing pixels = 50% | | | | | | | | | | |
| Method | *Butterfly* | *Bike* | *Castle* | *House* | *Girl* | *Light* | *Mickey* | *Parrots* | *Vegetable* | *Zebra* | Averaged |
| BPFA | 30.98/0.9595 | 29.01/0.9512 | 28.83/0.9486 | 39.12/0.9809 | 30.58/0.9598 | 25.36/0.9429 | 29.43/0.9501 | 33.97/0.9749 | 27.51/0.9468 | 26.35/0.9325 | 30.11/0.9547 |
| IPPO | 31.69/0.9724 | 29.90/0.9639 | 29.57/0.9576 | 40.03/0.9853 | 31.05/0.9672 | 26.76/0.9591 | 32.74/0.9719 | 36.19/0.9820 | 27.99/0.9536 | 28.42/0.9528 | 31.43/0.9666 |
| **Aloha** | 30.78/0.9414 | 29.78/0.9594 | 28.71/0.9485 | 40.56/0.9864 | 30.60/0.9608 | 25.83/0.9463 | 30.33/0.9515 | 36.14/0.9816 | 27.06/0.9475 | 27.70/0.9468 | 30.75/0.9570 |
| **JSM** | 31.47/0.9719 | 29.25/0.9618 | 29.48/0.9588 | 40.44/0.9853 | 30.59/0.9662 | 26.48/0.9594 | 30.69/0.9685 | 35.40/0.9805 | 27.72/0.9536 | 27.70/0.9545 | 30.92/0.9661 |
| **TNNM** | 32.09/0.9740 | 29.76/0.9640 | 29.63/0.9569 | 40.76/0.9861 | 30.89/0.9664 | 26.79/0.9601 | 32.23/0.9687 | 36.17/0.9811 | 27.64/0.9502 | 28.21/0.9569 | 31.42/0.9664 |
| **WNNM** | 33.17/0.9786 | 31.34/0.9751 | 29.74/0.9620 | 41.88/0.9893 | 32.09/0.9752 | 26.43/0.9620 | 33.35/0.9757 | 37.18/0.9852 | 28.40/0.9605 | 28.21/0.9631 | 32.18/0.9727 |
| **Proposed ($L_{1/2}$)** | 33.23/0.9788 | 31.41/0.9750 | 30.29/0.9637 | 41.80/0.9891 | 32.14/0.9750 | 27.53/0.9670 | 33.91/0.9769 | 37.87/0.9851 | 28.51/0.9608 | 29.32/0.9671 | 32.60/0.9739 |
| **Proposed ($L_{2/3}$)** | 33.00/0.9778 | 30.97/0.9729 | 30.25/0.9629 | 41.12/0.9873 | 31.93/0.9732 | 27.43/0.9661 | 33.59/0.9749 | 37.52/0.9837 | 28.29/0.9594 | 29.17/0.9652 | 32.33/0.9723 |

To further demonstrate the visual effect of our proposed nonconvex framework, Fig. 6 (c) to (j) present nine recovered images from 20% random samples of 'zebra' by our proposed method and other state-of-the-art competing methods, and (a) (b) is the original image and the damaged image for comparisons. It can be observed obviously that our proposed method can recover the corrupted image with higher quality and can recover more image details effectively.

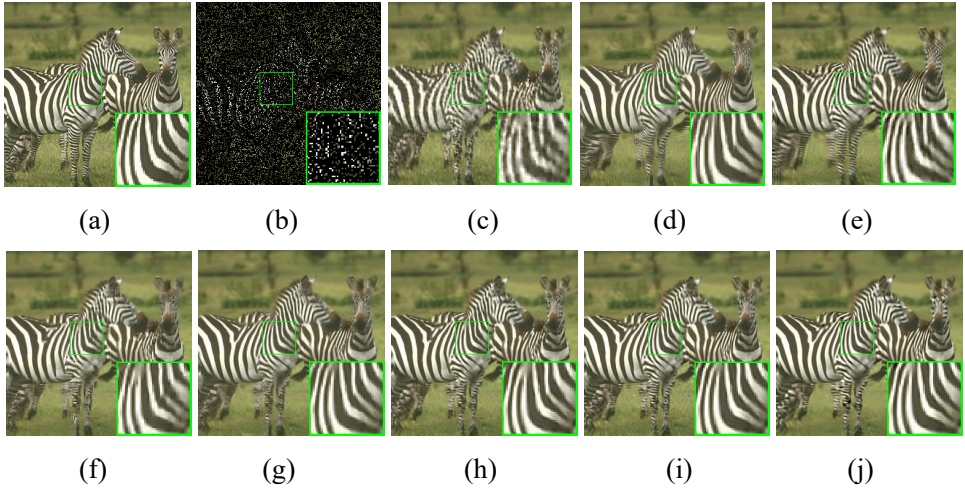

(a)　　　(b)　　　(c)　　　(d)　　　(e)

(f)　　　(g)　　　(h)　　　(i)　　　(j)

**Fig. 6**. Visual comparison of the original image, the damaged image, and nine reconstructed images by BPFA, IPPO, Aloha, JSM and our proposed algorithms for 80% missing of zebra. (a) Original Image; (b) The damaged image; (c) BPFA, PSNR=20.99 dB, FSIM=0.8190; (d) IPPO, PSNR=22.71 dB, FSIM=0.8665; (e) Aloha, PSNR=22.74 dB, FSIM=0.8565; (f) JSM, PSNR=21.86 dB, FSIM=0.8520; (g) TNNM, PSNR=22.68 dB, FSIM=0.8645; (h) WNNM, PSNR=22.85 dB, FSIM=0.8816; (i) Proposed ($p = 1/2$), PSNR=**23.24** dB, FSIM=**0. 8902**; (j) Proposed ($p = 2/3$), PSNR=22.63 dB, FSIM=**0. 8817**.

## II. Image inpainting for Text Removal

In this subsection, we will deal with another typical image inpainting problem for text removal, where the degraded matrix **A** is generated by a text mask. The goal of text removal is to reconstruct the original image from the degraded observations by removing the text. **Table 6** displays our results achieved by our proposed algorithm and other four state-of-the-art algorithms, our proposed approach can outperform other competitive methods significantly. To make a visual comparison, Fig. 7 (c) to (h) present the case of text removal from a corrupted 'Mickey' image, (a) and (b) denote the original image and the corrupted image for comparison. From the results we can find obviously that our proposed method can reconstruct image with more details and can remove more artifacts effectively.

**Table 6.** The PSNR (dB) values achieved by our proposed algorithm and four competing algorithms

| | | | | | Text Removal | | | | | | |
|---|---|---|---|---|---|---|---|---|---|---|---|
| **Method** | *Barbara* | *Butterfly* | *Castle* | *Clock* | *Cowboy* | *Girl* | *House* | *Mickey* | *Peppers* | *Starfish* | **Averaged** |
| **BPFA** | 34.13/0.9781 | 30.81/0.9580 | 30.25/0.9571 | 33.10/0.9676 | 30.43/0.9669 | 30.46/0.9618 | 37.65/0.9709 | 30.92/0.9607 | 36.04/0.9775 | 32.65/0.9695 | 32.64/0.9668 |
| **IPPO** | 37.65/0.9839 | 33.90/0.9760 | 31.91/0.9903 | 36.76/0.9861 | 32.62/0.9805 | 32.63/0.9794 | 41.20/0.9903 | 34.04/0.9838 | 39.51/0.9916 | 35.35/0.9859 | 35.56/0.9848 |
| **Aloha** | 39.16/0.9906 | 31.58/0.9569 | 30.34/0.9672 | 34.86/0.9770 | 30.94/0.9741 | 30.84/0.9693 | 41.37/0.9889 | 30.48/0.9641 | 37.40/0.9866 | 32.06/0.9719 | 33.90/0.9747 |
| **JSM** | 37.75/0.9888 | 33.05/0.9830 | 32.26/0.9763 | 35.86/0.9838 | 32.41/0.9801 | 32.14/0.9769 | 41.28/0.9891 | 32.93/0.9810 | 39.31/0.9911 | 35.18/0.9558 | 35.22/0.9806 |
| **TNNM** | 38.70/0.9908 | 33.39/0.9818 | 31.91/0.9746 | 36.55/0.9851 | 32.41/0.9788 | 32.32/0.9767 | 41.53/0.9903 | 33.03/0.9804 | 39.38/0.9911 | 34.70/0.9839 | 35.39/0.9834 |
| **WNNM** | 39.54/0.9919 | 34.12/0.9838 | 32.60/0.9778 | 37.36/0.9874 | 32.93/0.9809 | 32.81/0.9791 | 41.88/0.9907 | 33.83/0.9829 | 39.93/0.9921 | 35.44/0.9862 | 36.04/0.9853 |
| **Proposed ($L_{1/2}$)** | **40.53/0.9934** | 34.54/0.9849 | **32.90/0.9797** | 38.10/0.9895 | **33.25/0.9827** | 33.22/0.9811 | 42.65/0.9923 | 34.55/0.9853 | **40.51/0.9932** | 36.01/0.9880 | 36.63/0.9870 |
| **Proposed ($L_{2/3}$)** | **40.93/0.9938** | **34.67/0.9848** | 32.78/0.9804 | **38.29/0.9898** | 33.02/0.9833 | **33.10/0.9814** | **42.91/0.9925** | **34.56/0.9856** | 40.29/0.9931 | **36.20/0.9884** | **36.68/0.9873** |

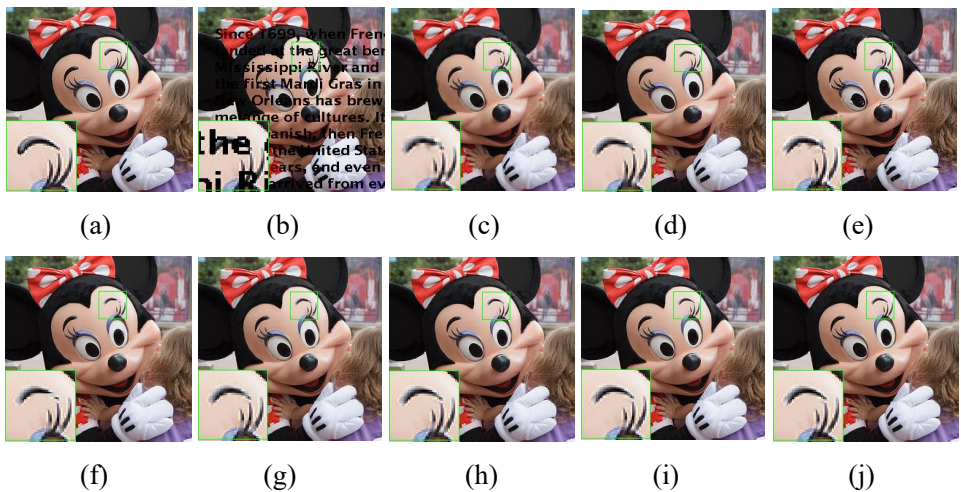

(a)      (b)      (c)      (d)      (e)

(f)      (g)      (h)      (i)      (j)

**Fig. 7**. Visual comparison of the original image, the corrupted image and nine reconstructed images by BPFA, IPPO, Aloha, JSM and our proposed algorithms for text remove of Mickey. (a) Original Image; (b) The corrupted image; (c) BPFA, PSNR=30.92 dB, FSIM=0.9607; (d) IPPO, PSNR=34.04 dB, FSIM=0.9838; (e) Aloha, PSNR=30.48 dB, FSIM=0.9641; (f) JSM, PSNR=32.93 dB, FSIM=0.9810; (g) TNNM, PSNR=33.03 dB, FSIM=0.9804; (h) WNNM, PSNR=33.83 dB, FSIM=0.9829; (i) Proposed ($p = 1/2$), PSNR=**34.55** dB, FSIM=**0.9853**; (j) Proposed ($p = 2/3$), PSNR=**34.56** dB, FSIM=**0.9856**.

### 4.3 Image deblurring

In this application, we focus on the case of image deblurring, where the blurred image is generated by the blur kernel and Gaussian noise with the standard deviation is σ. In this paper, three typical blur kernels are adopted in our experiments, the $9 \times 9$ Uniform blur kernel, the Gaussian blur kernel, and the Motion blur kernel. We compare our proposed nonconvex method to four recently approaches, i.e., MSEPLL [51], GSR-TNNM [33], JSM [50] and WNNM [6]. It should be noted that the method of GSR-TNNM is a recently proposed IR method, and the MSEPLL and JSM are two state-of-the art methods that can often achieve good results for image deblurring problem. The parameter selections for different typical blur kernels and functions are listed in the Table 7. The achieved PSNR and FSIM results on all test images are listed in the Table 8, and Fig. 8, 9 and 10 are presented the recovered under three blur kernels, we can find that our proposed method can produce the best results when the images are corrupted by the uniform blur kernel and the motion blur kernel. For the case of gaussian blur kernel, though our proposed method cannot achieve the best results in some cases, we still can achieve the higher average vales of PSNRs and FSIMs, and from Fig. 8, 9 and 10 we can see that our method can achieve much sharper image edges and cleaner textures than other methods.

**Table 7.** The parameter selection for different typical blur kernels and functions

| Different penalties | Uniform blur kernel | Gaussian blur kernel | Motion blur kernel |
|---|---|---|---|
| $p = \frac{1}{2}$ $(\mu, \lambda)$ | $(1e-2, 0.6)$ | $(8e-3, 0.13)$ | $(1e-2, 0.25)$ |
| $p = \frac{2}{3}$ $(\mu, \lambda)$ | $(8e-2, 0.006)$ | $(1.5e-2, 0.003)$ | $(1e-2, 0.005)$ |

**Table 8.** PSNR/FSIM comparisons for image deblurring

| $9 \times 9$ Uniform blur kernel, $\sigma = \sqrt{2}$ ||||||||||
|---|---|---|---|---|---|---|---|---|---|
| Method | Barbara | Bike | Fence | Flower | Parrots | Pepper | Starfish | Zebra | Averaged |
| Deblurred | 21.74/0.6720 | 19.56/0.6443 | 19.57/0.5949 | 23.53/0.6960 | 23.85/0.8265 | 24.03/0.7869 | 22.55/0.7439 | 17.82/0.5919 | 21.58/0.6946 |
| MSEPLL | 25.24/0.8791 | 24.15/0.8570 | 27.18/0.8990 | 28.09/0.8847 | 29.89/0.9246 | 31.23/0.9309 | 27.83/0.9045 | 23.05/0.8593 | 27.08/0.8924 |
| JSM | 25.68/0.8615 | 23.89/0.8368 | 27.25/0.9038 | 26.88/0.8420 | 27.87/0.8265 | 28.26/0.8458 | 26.63/0.8554 | 23.30/0.8292 | 26.22/0.8501 |
| GSR-TNNM | 26.68/0.8857 | 24.30/0.8375 | 28.38/0.9124 | 28.10/0.8697 | 29.59/0.9049 | 30.43/0.9003 | 27.71/0.8752 | 23.51/0.8265 | 27.34/0.8765 |
| WNNM | 27.59/0.9033 | 25.32/0.8637 | 29.56/0.9302 | 28.88/0.8856 | 30.51/0.9088 | 31.21/0.9121 | 28.70/0.8989 | **24.71/0.8678** | 28.31/0.8963 |
| Proposed ($L_{1/2}$) | 27.64/0.9053 | 25.34/0.8645 | 29.67/0.9309 | 28.96/0.8880 | 30.65/0.9154 | 31.38/0.9168 | **28.78/0.9042** | 23.69/0.8484 | 28.26/**0.8967** |
| Proposed ($L_{2/3}$) | 27.40/0.9091 | 25.07/0.8609 | 29.76/0.9285 | 29.21/0.8932 | 30.87/0.9359 | 32.14/0.9332 | 28.75/0.9027 | 24.30/0.8651 | **28.44/0.9036** |

| Gaussian blur kernel: fspecial ("gaussian", 25, 1.6) $\sigma = \sqrt{2}$ ||||||||||
|---|---|---|---|---|---|---|---|---|---|
| Method | Barbara | Bike | Fence | Flower | Parrots | Pepper | Starfish | Zebra | Averaged |
| Deblurred | 22.79/0.7568 | 22.19/0.7708 | 22.14/0.6830 | 26.57/0.8285 | 26.91/0.8937 | 27.79/0.8876 | 25.79/0.8506 | 20.20/0.7113 | 24.30/0.7978 |
| MSEPLL | 23.83/0.8610 | 26.00/**0.9021** | 25.97/0.8936 | 29.98/**0.9200** | 31.49/**0.9460** | **33.28/0.9530** | 29.98/0.9362 | 23.64/0.8857 | 28.02/0.9122 |
| JSM | 25.78/0.8738 | 26.65/0.8890 | 27.08/0.9028 | 30.01/0.8956 | 31.07/0.8928 | 31.98/0.9097 | 30.08/0.9122 | 24.66/0.8726 | **28.41/0.8936** |
| GSR-TNNM | **26.76/0.9012** | 26.64/0.8854 | 27.58/0.9128 | 30.50/0.9108 | 31.91/0.9338 | 32.75/0.9279 | 30.42/0.9191 | 24.78/0.8711 | 28.92/0.9078 |
| WNNM | 26.55/0.8826 | 26.50/0.8785 | 27.31/0.9065 | 29.71/0.8855 | 30.96/0.8831 | 31.40/0.8916 | 29.74/0.8965 | 24.73/0.8634 | 28.36/0.8860 |
| Proposed ($L_{1/2}$) | 26.75/0.8952 | 26.65/0.8844 | 27.52/0.9111 | 30.32/0.9045 | 31.59/0.9148 | 32.32/0.9158 | 30.89/0.9301 | 24.86/0.8809 | 28.86/0.9046 |
| Proposed ($L_{2/3}$) | 26.73/0.9006 | 26.85/0.8948 | 27.64/0.9150 | 30.71/0.9154 | 32.13/0.9392 | 33.21/0.9385 | 30.72/0.9258 | 24.87/0.8786 | 29.11/0.9135 |

| Motion blur kernel: fspecial ("motion", 20, 45) $\sigma = \sqrt{2}$ ||||||||||
|---|---|---|---|---|---|---|---|---|---|
| Method | Barbara | Bike | Fence | Flower | Parrots | Pepper | Starfish | Zebra | Averaged |
| Deblurred | 20.92/0.6678 | 17.99/0.6303 | 19.65/0.5641 | 22.35/0.6966 | 22.48/0.8201 | 22.17/0.7656 | 21.01/0.7520 | 16.96/0.6009 | 20.44/0.6872 |
| MSEPLL | 25.98/0.8944 | 25.54/0.8924 | 25.64/0.8763 | 28.99/0.9022 | 31.49/0.9318 | 30.67/0.9283 | 27.81/0.9142 | 24.42/0.8850 | 27.57/0.9031 |
| JSM | 24.95/0.8464 | 23.19/0.8342 | 25.40/0.8809 | 25.69/0.8175 | 26.48/0.7878 | 26.38/0.8071 | 25.14/0.8288 | 22.60/0.8201 | 24.98/0.8279 |
| GSR-TNNM | 28.45/0.9072 | 25.76/0.8688 | 27.55/0.9030 | 29.63/0.8956 | 31.56/0.9197 | 30.96/0.9074 | 28.28/0.8913 | 24.56/0.8515 | 28.34/0.8931 |
| WNNM | 29.90/0.9347 | 27.18/0.9036 | 28.81/0.9163 | 30.59/0.9145 | 33.16/0.9471 | 33.00/0.9457 | 30.16/0.9286 | 25.61/0.8926 | 29.80/0.9229 |
| Proposed ($L_{1/2}$) | 30.14/0.9371 | 27.40/0.9070 | 29.05/0.9226 | 30.71/0.9167 | 33.13/0.9452 | 33.03/0.9448 | 30.34/0.9300 | 25.88/0.8983 | 29.96/0.9252 |
| Proposed ($L_{2/3}$) | 30.38/0.9397 | 27.48/0.9070 | 29.11/0.9228 | 30.78/0.9171 | 33.20/0.9463 | 33.15/0.9463 | 30.48/0.9309 | 26.02/0.9001 | 30.08/0.9263 |

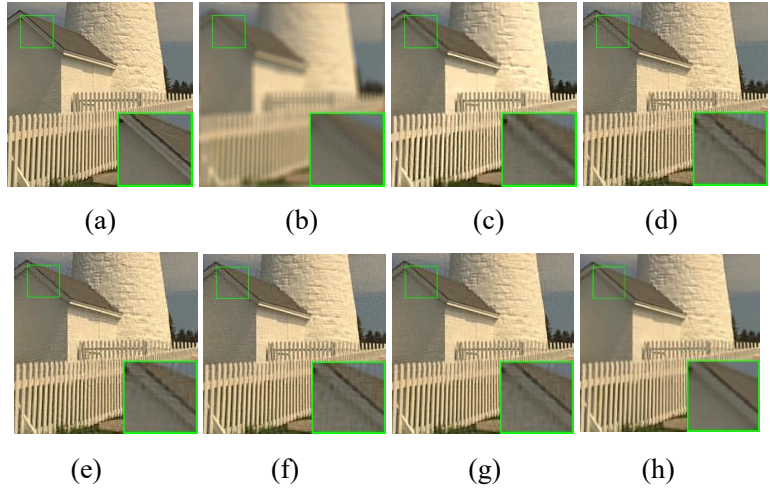

**Fig. 8**. Restored *Fence* images form corrupted image by $9 \times 9$ Uniform blur kernel with $\sigma = \sqrt{2}$ by various methods for visual comparison. (a) Original Image; (b) The deblurred and noisy image, PSNR=19.57 dB, FSIM=0.5949; (c) MSEPLL, PSNR=27.18 dB, FSIM=0.8990; (d) JSM, PSNR=27.25 dB, FSIM=0.9038; (e) GSR-TNNM, 28.38 dB, FSIM=0.9124; (f) WNNM, PSNR=29.56 dB, FSIM=0.9302; (g) Proposed ($p = 1/2$), PSNR=**29.67** dB, FSIM=**0.9309**; (h) Proposed ($p = 2/3$), PSNR=**29.76** dB, FSIM=**0.9285**.

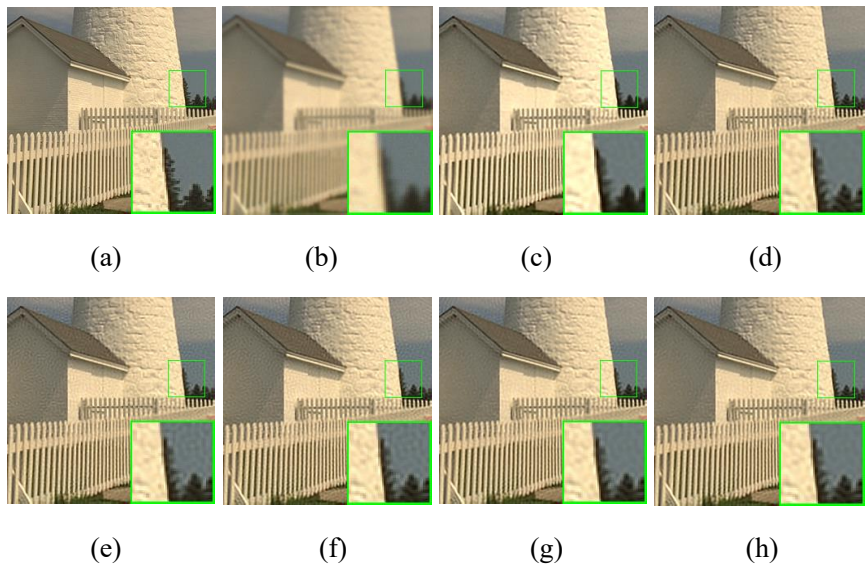

**Fig. 9**. Restored *Fence* images form corrupted image by Gaussian blur kernel with $\sigma = \sqrt{2}$ by various methods for visual comparison. (a) Original Image; (b) The deblurred and noisy image, PSNR=22.14 dB, FSIM=0.6830; (c) MSEPLL, PSNR=25.97 dB, FSIM=0.8936; (d) JSM, PSNR=27.08 dB, FSIM=0.9028; (e) GSR-TNNM, 27.58 dB, FSIM=0.9128; (f) WNNM, PSNR=27.31 dB, FSIM=0.9065; (g) Proposed $p = 1/2$), PSNR=27.52 dB, FSIM=0.9111; (h) Proposed ($p = 2/3$), PSNR=**27.64** dB, FSIM=**0.9150**.

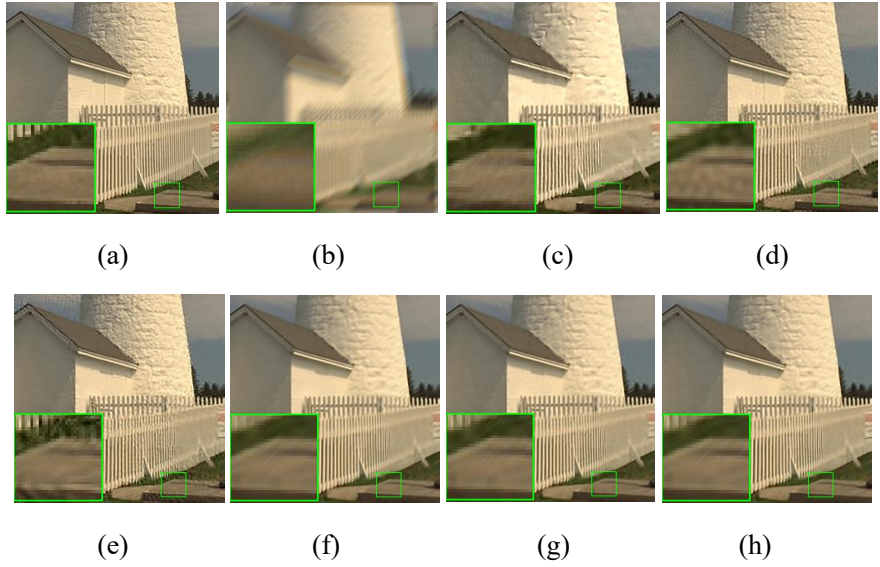

(a)              (b)              (c)              (d)

(e)              (f)              (g)              (h)

**Fig. 10**. Visual comparison of the original image, the deblurred and noisy image corrupted by motion blur kernel with $\sigma = \sqrt{2}$. (a) Original Image; (b) The deblurred and noisy image, PSNR=19.65 dB, FSIM=0.5641; (c) MSEPLL, PSNR=25.64 dB, FSIM=0.8763; (d) JSM, PSNR=25.40 dB, FSIM=0.8809; (e) GSR-TNNM, 27.55 dB, FSIM=0.9030; (f) WNNM, 28.81 dB, FSIM=0.9163; (g) Proposed ($p = 1/2$), PSNR=**29.05** dB, FSIM=**0.9226**; (h) Proposed ($p = 2/3$), PSNR=**29.11** dB, FSIM=**0.9228**.

### 4.4 Salt and Pepper Noise Removal

As a typical impulsive noise, the salt and pepper noise (SPN) usually occurs in the procedure of the image acquisition and transmission. In this application, we will adopt our proposed algorithm to recover the corrupted image by SPN with different strengths, here, five state-of-the art algorithms of WESNR [52], JSM [50], WCSR [53], TNNM [33] and WNNM [6] are compared with our method. To handle this problem, an adaptive median filter [54] is first applied to the corrupted images, hence we can identify the maks matrix **A**, and we can transform the noise removal problem into the IR problem. Table 9 details the parameter selections for different penalties and the noise density. Table 10 presents all the PSNR/FSIM results on six test images, we can find that our proposed method can outperform other three methods significantly. Moreover, some visual results of 'Cameraman' for five algorithms are presented in the fig. 11, by comparing with the competing results in (c), (d) and (e), we can obviously find that our method can earn the most visually results, e.g., (f) and (g).

**Table 9.** The parameter selection for different sub-rate and functions

| Different penalties | 30% SPN | 50% SPN | 80% SPN |
|---|---|---|---|
| $p = \frac{1}{2}$ $(\mu, \lambda)$ | $(1e-2, 0.15)$ | $(5e-3, 0.1)$ | $(5e-3, 0.4)$ |
| $p = \frac{2}{3}$ $(\mu, \lambda)$ | $(1e-1, 0.05)$ | $(1e-2, 0.01)$ | $(2e-2, 0.03)$ |

**Table 10.** PSNR/FSIM comparison for images corrupted by salt and pepper noise

| 30% SPN | | | | | | | |
|---|---|---|---|---|---|---|---|
| Method | *Cameraman* | *Couple512* | *Hill512* | *House* | *Man512* | *Straw* | Averaged |
| **Noisy** | 10.33/0.4542 | 10.76/0.6369 | 10.61/0.5731 | 10.75/0.4081 | 10.64/0.5864 | 10.64/0.6698 | 10.63/0.5548 |
| **WESNR** | 28.15/0.9303 | 32.43/0.9800 | 33.16/0.9827 | 36.26/0.9548 | 32.99/0.9816 | 28.24/0.9442 | 31.87/0.9623 |
| JSM | 30.47/0.9132 | 32.22/0.9751 | 31.85/0.9654 | 35.58/0.9249 | 31.77/0.9674 | 27.93/0.9370 | 31.64/0.9472 |
| WCSR | 32.32/0.9704 | 35.95/0.9940 | 36.81/0.9939 | 39.70/0.9861 | 35.86/0.9935 | 31.36/0.9769 | 35.33/0.9858 |
| TNNM | 31.86/0.9652 | 35.08/0.9913 | 35.76/0.9905 | 40.73/0.9842 | 35.10/0.9908 | 30.88/0.9720 | 34.90/0.9823 |
| WNNM | 33.23/0.9763 | 36.77/0.9946 | 37.19/0.9941 | 42.91/0.9917 | 36.66/0.9950 | 33.23/0.9843 | 36.67/0.9893 |
| Proposed ($L_{1/2}$) | **33.24/0.9767** | **36.85/0.9946** | **37.30/0.9943** | **43.11/0.9920** | **36.79/0.9948** | **33.37/0.9848** | **36.78/0.9895** |
| Proposed ($L_{2/3}$) | **33.07/0.9749** | **36.88/0.9946** | **37.31/0.9940** | **42.79/0.9903** | **36.81/0.9944** | **33.52/0.9853** | **36.73/0.9889** |
| 50% SPN | | | | | | | |
| Method | *Cameraman* | *Couple512* | *Hill512* | *House* | *Man512* | *Straw* | *Averaged* |
| Noisy | 8.10/0.3610 | 8.53/0.5546 | 8.39/0.4977 | 8.53/0.4081 | 8.42/0.5864 | 8.43/0.5931 | 8.40/0.5002 |
| WESNR | 25.29/0.8901 | 30.40/0.9665 | 31.32/0.9675 | 34.39/0.3188 | 30.92/0.5075 | 25.77/0.9039 | 29.68/0.7591 |
| JSM | 28.60/0.8865 | 30.79/0.9635 | 30.78/0.9519 | 34.51/0.9140 | 30.49/0.9522 | 25.98/0.9023 | 30.19/0.9284 |
| WCSR | 28.98/0.9383 | 32.47/0.9842 | 33.51/0.9852 | 35.75/0.9681 | 32.51/0.9837 | 27.68/0.9479 | 31.82/0.9679 |
| TNNM | 29.32/0.9473 | 32.85/0.9853 | 33.74/0.9848 | 38.44/0.9800 | 33.05/0.9853 | 27.97/0.9466 | 32.56/0.9716 |
| WNNM | 30.39/0.9570 | 33.80/0.9882 | 34.40/0.9870 | 39.29/0.9829 | 33.75/0.9879 | 29.75/0.9637 | 33.56/0.9778 |
| Proposed ($L_{1/2}$) | **30.65/0.9606** | **34.27/0.9897** | **34.92/0.9887** | **40.08/0.9855** | **34.24/0.9898** | **30.35/0.9682** | **34.09/0.9804** |
| Proposed ($L_{2/3}$) | **30.66/0.9601** | **34.25/0.9896** | **34.86/0.9883** | **39.87/0.9841** | **34.09/0.9893** | **30.36/0.9682** | **34.01/0.9799** |
| 80% SPN | | | | | | | |
| Method | *Cameraman* | *Couple512* | *Hill512* | *House* | *Man512* | *Straw* | *Averaged* |
| Noisy | 6.04/0.3012 | 6.48/0.4898 | 6.33/0.4354 | 6.44/0.2651 | 6.37/0.4422 | 6.36/0.5244 | 6.34/0.4097 |
| WESNR | 20.32/0.7907 | 25.91/0.9058 | 27.93/0.9210 | 27.12/0.8595 | 27.09/0.9194 | 20.96/0.7439 | 24.89/0.8567 |
| JSM | 24.18/0.7949 | 26.95/0.9005 | 27.83/0.8901 | 30.98/0.8737 | 27.14/0.8811 | 20.90/0.6971 | 26.33/0.8396 |
| WCSR | 23.15/0.8340 | 26.45/0.9302 | 28.26/0.9435 | 28.60/0.8906 | 27.26/0.9366 | 21.52/0.8250 | 25.87/0.8933 |
| TNNM | 24.40/0.8458 | 26.79/0.9137 | 28.10/0.9176 | 31.61/0.9210 | 27.57/0.9157 | 21.09/0.7606 | 26.59/0.8791 |
| WNNM | 25.58/0.8832 | 28.77/0.9524 | 29.65/0.9564 | 34.01/0.9466 | 29.07/0.9494 | 24.12/0.8766 | 28.53/0.9274 |
| Proposed ($L_{1/2}$) | **25.62/0.8866** | **28.85/0.9543** | **29.77/0.9526** | **34.10/0.9481** | **29.13/0.9512** | **24.32/0.8840** | **28.63/0.9295** |
| Proposed ($L_{2/3}$) | **25.64/0.8876** | **28.86/0.9551** | **29.91/0.9525** | **34.05/0.9451** | **29.12/0.9510** | **24.48/0.8914** | **28.68/0.9305** |

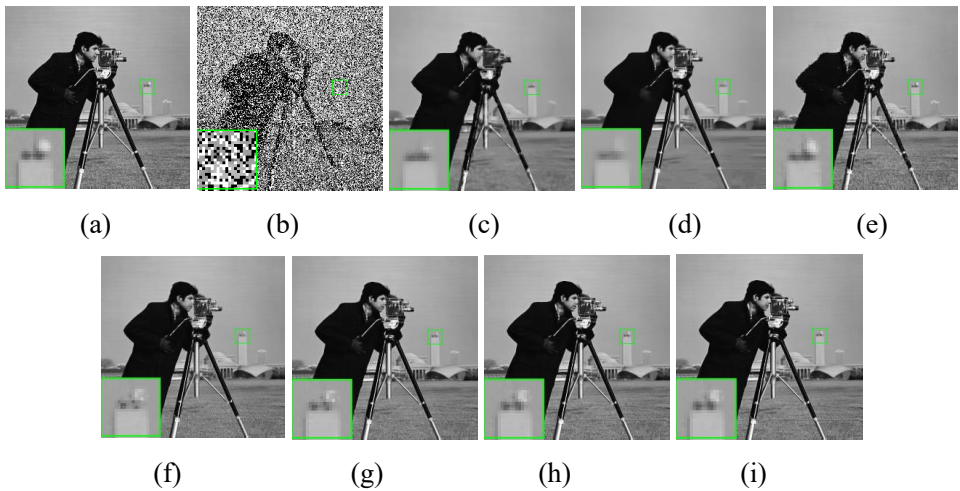

(a)      (b)      (c)      (d)      (e)

(f)      (g)      (h)      (i)

**Fig. 11**. Visual comparison of the original image, the corrupted image by 50% salt and pepper noise, and five reconstructed images by MSEPLL, GSR-TNNM, JSM and our proposed algorithms. (a) Original Image; (b) The noisy image; (c) WESNR, PSNR=25.29 dB, FSIM=0.8901; (d) JSM, 28.60 dB, FSIM=0.8865; (e) WCSR, PSNR=28.98 dB, FSIM=0.9383; (f) TNNM, PSNR=29.32 dB,

FSIM=0.9473; (g) WNNM, PSNR=30.39 dB, FSIM=0.9570; (h) Proposed ( $p = 1/2$ ), PSNR=**30.65** dB, FSIM=**0.9606**; (i) Proposed ($p = 2/3$), PSNR=**30.66** dB, FSIM=**0.9601**.

## 4.5 Convergence analysis

Although our proposed method can achieve very effective results, it is intractable to give a theoretical proof of the convergence for our proposed algorithm since the relaxation function of nuclear norm is nonconvex. It is well documented that the behavior of PSNR curves versus the iteration number can reflect the convergence property visually [21][31]. **Fig. 12** (a), (b), (c) and (d) present the PSNRs curves for different IR problems, including the image compressive sensing problem under different sub-sampling rates, image inpainting problem from partial random samples, image inpainting for text removal, and SPN removal. It is obviously that our proposed algorithm contains good convergence property and robustness.

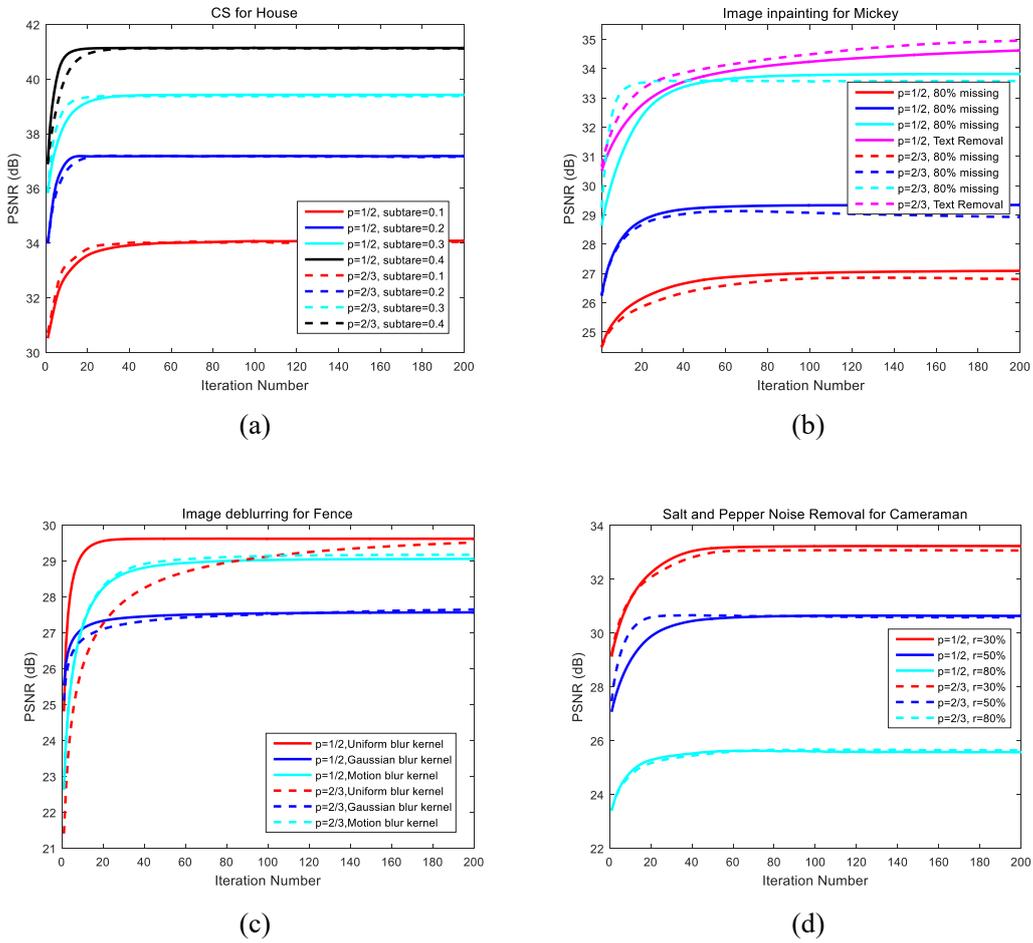

**Fig. 12**. The verification of the convergence of our proposed method. (a) Image compressive sensing; (b) Image inpainting; (c) Image deblurring; (d) Salt and pepper noise removal.

## V. Conclusion

This paper proposed a novel GSC based denoising model via rank minimization for IR applications. We first converted the group sparse coding problem into the low-rank minimization problem via an effective adaptive dictionary learning strategy. To better approximate the rank of the group matrix, we developed a novel DNNR minimization framework for GSC. To solve this nonconvex optimization problem, an GSC based iteratively-reweighted singular value function thresholding algorithm was proposed. Finally, some typical IR applications have been considered to evaluate the effectiveness and priority of our proposed method via the ADMM strategy.

## Acknowledgement

This work was supported by the Project Funded by the National Science and Technology Major Project of the Ministry of Science and Technology of China under Grant TC190A3WZ-2, National Natural Science Foundation of China under Grant 61901228 and 61671253, the Innovation and Entrepreneurship of Jiangsu High-level Talent under Grant CZ0010617002, the Six Top Talents Program of Jiangsu under Grant XYDXX-010, the 1311 Talent Plan of Nanjing University of Posts and Telecommunications.